\documentclass[12pt]{iopart}

\usepackage{hyperref} 
\usepackage{graphicx}
\usepackage{color}
\usepackage{dcolumn}
\usepackage{bm}
\usepackage{verbatim} 
\usepackage{cite} 
\usepackage[normalem]{ulem}

\begin{document}


    \long\def \cblu#1{\color{blue}#1}
    \long\def \cred#1{\color{red}#1}
    \long\def \cgre#1{\color{green}#1}
    \long\def \cpur#1{\color{purple}#1}

\newcommand{\eric}[1]{{\color{blue}#1}}
\newcommand{\guido}[1]{{\color{violet}#1}}
\newcommand{\matthias}[1]{{\color{blue}#1}}
\newcommand{\fabian}[1]{{\color{blue}#1}}
\newcommand{\di}[1]{{\color{blue}#1}}
\newcommand{\ericC}[1]{{\color{red}\textit{\textbf{Eric:} #1}}}
\newcommand{\guidoC}[1]{{\color{red}\textit{\textbf{Guido:} #1}}}
\newcommand{\matthiasC}[1]{{\color{red}\textit{\textbf{Matthias:} #1}}}
\newcommand{\fabianC}[1]{{\color{red}\textit{\textbf{Fabian:} #1}}}
\newcommand{\diC}[1]{{\color{red}\textit{\textbf{Di:} #1}}}

\def\FileRef{
\input FName
{
\newcount\hours
\newcount\minutes
\newcount\min
\hours=\time
\divide\hours by 60
\min=\hours
\multiply\min by 60
\minutes=\time
\
\advance\minutes by -\min
{\small\rm\em\the\month/\the\day/\the\year\ \the\hours:\the\minutes
\hskip0.125in{\tt\FName}
}
}}

\mathchardef\muchg="321D
\let\na=\nabla
\let\pa=\partial

\let\muchg=\gg

\let\t=\tilde
\let\ga=\alpha
\let\gb=\beta
\let\gc=\chi
\let\gd=\delta
\let\gD=\Delta
\let\ge=\epsilon
\let\gf=\varphi
\let\gg=\gamma
\let\gh=\eta
\let\gj=\phi
\let\gF=\Phi
\let\gk=\kappa
\let\gl=\lambda
\let\gL=\Lambda
\let\gm=\mu
\let\gn=\nu
\let\gp=\pi
\let\gq=\theta
\let\gr=\rho
\let\gs=\sigma
\let\gt=\tau
\let\gw=\omega
\let\gW=\Omega
\let\gx=\xi
\let\gy=\psi
\let\gY=\Psi
\let\gz=\zeta

\let\lbq=\label
\let\rfq=\ref
\let\na=\nabla
\def\daI{{\dot{I}}}
\def\daX{{\dot{X}}}
\def\dsq{{\dot{q}}}
\def\dgj{{\dot{\phi}}}

\def\bgs{\bar{\sigma}}
\def\bgh{\bar{\eta}}
\def\bgg{\bar{\gamma}}
\def\bgy{\bar{\psi}}
\def\bgF{\bar{\Phi}}
\def\bgY{\bar{\Psi}}

\def\baF{\bar{F}}
\def\baG{\bar{G}}
\def\bsj{\bar{j}}
\def\baJ{\bar{J}}
\def\bsp{\bar{p}}
\def\baP{\bar{P}}
\def\bsx{\bar{x}}

\def\hgj{\hat{\phi}}
\def\hgq{\hat{\theta}}

\def\HaT{\hat{T}}
\def\HaR{\hat{R}}
\def\Hsb{\hat{b}}
\def\Hsh{\hat{h}}
\def\Hsz{\hat{z}}

\let\gG=\Gamma
\def\taA{{\tilde{A}}}
\def\taB{{\tilde{B}}}
\def\taG{{\tilde{G}}}
\def\tsp{{\tilde{p}}}
\def\tsv{{\tilde{v}}}
\def\tgF{{\tilde{\Phi}}}

\def\wgx{{\bm{\xi}}}
\def\wgz{{\bm{\zeta}}}

\def\wsb{{\bf b}}
\def\wse{{\bf e}}
\def\wsk{{\bf k}}
\def\wsi{{\bf i}}
\def\wsj{{\bf j}}
\def\wsl{{\bf l}}
\def\wsn{{\bf n}}
\def\wsp{{\bf p}}
\def\wsr{{\bf r}}
\def\wsu{{\bf u}}
\def\wsv{{\bf v}}
\def\wsx{{\bf x}}

\def\vaB{\vec{B}}
\def\vse{\vec{e}}
\def\vsh{\vec{h}}
\def\vsl{\vec{l}}
\def\vsv{\vec{v}}
\def\vgn{\vec{\nu}}
\def\vgk{\vec{\kappa}}
\def\vgt{\vec{\gt}}
\def\vgx{\vec{\xi}}
\def\vgz{\vec{\zeta}}

\def\waA{{\bf A}}
\def\waB{{\bf B}}
\def\waD{{\bf D}}
\def\waE{{\bf E}}
\def\waF{{\bf F}}
\def\waJ{{\bf J}}
\def\waV{{\bf V}}
\def\waX{{\bf X}}

\def\R#1#2{\frac{#1}{#2}}
\def\btbl{\begin{tabular}}
\def\etbl{\end{tabular}}
\def\bqbl{\begin{eqnarray}}
\def\eqbl{\end{eqnarray}}
\def\ebox#1{
  \begin{eqnarray}
    #1
\end{eqnarray}}


\def \cred#1{{\color{red}\sout{(#1)}}}
\def \cblu#1{{\color{blue}#1}}

\title[Runaway electron equilibrium solver]{Drift surface solver for runaway electron current dominant equilibria during the Current Quench}
\author{Lu Yuan$^{1}$ and Di Hu$^{1,*}$}
\address{
$^1$Beihang University, No. 37 Xueyuan Road, Haidian District, 100191 Beijing, China.
}
\address{
$^*$Corresponding author.
}
\ead{hudi2@buaa.edu.cn}

\vspace{10pt}
\begin{indented}
\item[]\today
\end{indented}

\begin{abstract}
Runaway electron current generated during the Current Quench phase of tokamak disruptions could result in severe damage to future high performance devices. To control and mitigate such runaway electron current, it is important to accurately describe the runaway electron current dominated equilibrium, based on which further stability analysis could be carried out. In this paper, we derive a Grad-Shafranov-like equation solving for the axisymmetric drift surfaces of the runaway electrons instead of the magnetic flux surfaces for the simple case that all runaway electron share the same parallel momentum. This new equilibrium equation is then numerically solved with simple rectangular wall with ITER-like and MAST-like geometry parameters. The deviation between the drift surfaces and the flux surfaces is readily obtained, and runaway electrons is found to be well confined even in regions with open field lines. The change of the runaway electron parallel momentum is found to result in a horizontal current center displacement without any changes in the total current or the external field. The runaway current density profile is found to affect the susceptibility of such displacement, with flatter profiles result in more displacement by the same momentum change. With up-down asymmetry in the external poloidal field, such displacement is accompanied by a vertical displacement of runaway electron current. It is found that this effect is more pronounced in smaller, compact device and weaker poloidal field cases. The above results demonstrate the dynamics of current center displacement caused by the momentum space change in the runaway electrons, and pave way for future, more sophisticated runaway current equilibrium theory with more realistic consideration on the runaway electron momentum distribution. This new equilibrium theory also provides foundation for future stability analysis of the runaway electron current.
\end{abstract}

%
%
%
\maketitle
%
%
\section{Introduction}
\label{s:Intro}
Plasma disruptions are the result of non-linearly unstable magneto-hydrodynamic (MHD) instabilities which destroy the global plasma confinement and result in rapid and localized energy deposition upon the Plasma Facing Components (PFCs), potentially causing irreversible damage to future high performance devices \cite{Lehnen2015JNM}. Among the threats posed by disruptions, the runaway electrons and their localized deposition onto the first wall are one of the most feared consequences \cite{Lehnen2015JNM,Nygren1997JNM,Gilligan1990JNM}  . In the most disastrous scenario, significant portion of the original plasma current could be converted into runaway electron current, and all of their kinetic energy as well as a large portion of the magnetic energy it carries could be unleashed locally \cite{Martin-Solis2017NF}. This should be avoided at all costs, and there have been a lot of effort going on regarding the control and depletion of such runaway current \cite{Wei Yunong2020NST,Paz-Soldan2019PPCF,Papp2011PPCF,Hollmann2015POP,Gobbin2018PPCF}.

To address the confinement and loss of runaway electrons in the configuration space, however, one should have a firm grasp on the equilibrium solution, upon which instability studies could be carried out.
The runaway electrons, on the other hand, will not exactly follow the magnetic field lines due to their high momentum. Once they become the main current carrier, the current density stops following the magnetic surface, and the current density (or it multiplied by the major radius in toroidal geometry) is no longer a flux function. Instead, the current will follow the drift surface of the runaway electrons, which is essentially the constant canonical angular momentum surface in the axisymmetric geometry \cite{Cary2009RMP,Chang Liu2018NF}. Thus the conventional Grad-Shafranov equation \cite{LZBook1986} for solving plasma equilibrium need to be revised to describe this new class of equilibria, and a new multi-fluid-like equilibrium equation \cite{Galeotti2011POP,Goedbloed2005POP,Sudan1979POF} needs to be derived which solves the runaway electron drift surfaces directly.

We are satisfied with the consideration of a simplified runaway current model in this study. We obtain the force balance equation for the strongly passing runaway electron guiding center from its Lagrangian \cite{Cary2009RMP}. Such force balance equation is then combined with the toroidal symmetry to yield a Grad-Shafranov-like second order derivative equation for the effective flux (the canonical angular momentum), which can then be solved with given boundary condition. For simplicity, we have assumed that all runaway electrons have the same parallel momentum, although they can have a spatial distribution in density. We then solve the equation numerically assuming static magnetic flux boundary condition for a range of geometric parameters, runaway electron currents, runaway electron momentum and poloidal magnetic fields. Apparent deviation could be found to exist between the calculated drift surfaces and the flux surfaces, depending on the runaway electron momentum, device geometry and the poloidal field strength. It is found that the current center will exhibit a horizontal displacement as the runaway electron momentum changes, even with constant total current and poloidal field, in consistent with results from previous simpler model \cite{Di2016POP}. The flattened current profile is more susceptible to such displacement with the same momentum change. Furthermore, in the presence of up-down asymmetry in the poloidal field, the runaway current exhibits vertical displacement accompanied by said horizontal displacement. Such displacement is purely caused by their parallel momentum change, and may have important implications to the stability of the Current Quench plasmas where runaway current is dominant, as well as to the runaway electron scrapping-off as a result of said displacement.

The rest of this paper is arranged as the following. In section \ref{s:EquilibriumEq}, we derive the new Grad-Shafranov-like equation for the runaway electrons from their guiding center Lagrangian. In Section \ref{s:NumericalEq}, we performed a numerical parameter scan of numerical solutions to investigate the equilibrium evolution under different geometric parameters, runaway electron current, parallel momentum and poloidal field. Finally, the discussion and conclusion is shown in Section \ref{s:Conclusion}.

\section{The equilibrium equation for runaway electron current}
\label{s:EquilibriumEq}

The Grad-Shafranov-like equation could be obtained from two ingredients: the force balance equation and a symmetric direction. For the latter, we are considering a simple 2D equilibrium in this paper, and the symmetric direction is simply the toroidal direction. The force balance equation for the cold Current Quench (CQ) plasma should simply be $\waJ\times\waB=0$ in the limit that the current carrier does not have strong parallel momentum. However, for runaway electrons, the centrifugal force as a result of their high momentum must be considered, and this is balanced by the $\waJ\times\waB$ term.

A simple way of obtaining such force balance equation is to look at the description of the guide center trajectory from its Lagrangian \cite{Cary2009RMP}, and realize that such trajectory must automatically satisfy the force balance for the individual particle. In principle, one should then construct a series of moment equations for the runaway electron Boltzmann equation to render the runaway electron ``fluid equations'' \cite{FFChenBook}, from which the force balance equation is obtained. However, we will only consider a simplified case here as a first step that all runaway electrons share the same parallel momentum ($\gd$ function distribution in momentum space for a given configuration space coordinate). In such case, the runaway current density could be expressed as
\bqbl
\lbq{eq:waJwsu}
\waJ
=
n_{RE}q\wsu
.\eqbl
Here $n_{RE}$ is the runaway electron density, $q$ is its charge and $\wsu$ is the same with the individual guiding center velocity.

The Hamiltonian equation of motion for each coordinate of the runaway electron guiding center are \cite{Cary2009RMP,Chang Liu2018NF}:
\bqbl
\lbq{eq:Xdot}
\dot{\waX}
=
\R{p_\|}{\gg m}\R{\waB^*}{B_\|^*}
+\R{\wsb}{qB_\|^*}\times\left(\na\gF^*+\R{\pa\waA^*}{\pa t}\right)
,\eqbl
\bqbl
\lbq{eq:p_paradot}
\dot{p_\|}
=
-\R{\waB^*}{B_\|^*}\cdot\left(\na\gF^*+\R{\pa\waA^*}{\pa t}\right)
,\eqbl
\bqbl
\lbq{eq:mudot}
\dot{\gm}
=
0
,\eqbl
\bqbl
\lbq{eq:thetadot}
\dot{\gq}
=
\R{qB}{\gg m}
.\eqbl
Here, $\waX$ is the configuration space coordinates, $p_\|$ is the parallel momentum, $\gm$ is the magnetic moment and $\gq$ is the gyro-angle, $\wsb$ is the magnetic field direction, $\gg$ is the relativistic factor, $q$ and $m$ are the electron charge and rest mass respectively. We have also defined the following effective fields:
\bqbl
\lbq{eq:defAStar}
\waA^*
\equiv
\waA+\R{p_\|}{q}\wsb
,\eqbl
\bqbl
\lbq{eq:defBStar}
\waB^*
\equiv
\na\times\waA^*
,\eqbl
\bqbl
B_\|^*
\equiv
\waB^*\cdot\wsb
,\eqbl
\bqbl
\na \gF^*
\equiv
\R{\gm\na B}{\gg}+q\na\gF
.\eqbl
Here $\waA$ is the vector potential and $\gF$ is the electric potential. 
It should be noted that, if we are to consider higher parallel momentum scenario where $p_\|/qBR\propto \ge$ with $\ge\equiv a/R$ being the inverse aspect ratio, as is assumed in Ref.\,\cite{Chang Liu2018NF}, then additional curvature drift terms needs to be added in the above equations.
In this study, however, we consider $p_\|/qBR\propto \ge^2$, so that the higher order curvature terms in Ref.\,\cite{Chang Liu2018NF} is considered as next order effects in this paper and neglected.
This ordering, although less ambitious, should already cover the major parameter range of interest during the Current Quench for strong magnetic field device such as ITER, since the radiative and collisional drag, as well as kinetic instabilities would tend to limit the upper bound of the runaway electrons' parallel momentum.
For strongly passing runaway electrons in quasi-neutral plasmas, we approximate $\na\gF^*\simeq0$. In steady states, all partial time derivative terms also vanish. So that we have
\bqbl
\lbq{eq:Xdot2}
\dot{\waX}
=
\R{p_\|}{\gg m}\R{\waB^*}{B_\|^*}
,\eqbl
\bqbl
\lbq{eq:p_paradot2}
\dot{p_\|}
=
0
.\eqbl
The parallel momentum is thus a constant along the trajectory line, thus a function of the drift surface, and the runaway electron trajectory is determined by the effective magnetic field $\waB^*$. Substituting into Eq.\,(\rfq{eq:waJwsu}), we obtain:
\bqbl
\lbq{eq:waJBstar}
\waJ
=
n_{RE}q\R{p_\|}{\gg m}\R{\waB^*}{B_\|^*}
.\eqbl
Note that Eq.\,(\rfq{eq:waJBstar}) automatically satisfy the force balance in the cases we consider here, and the force balance equation simply yields:
\bqbl
\lbq{eq:RE_ForceBalance}
\waJ\times\waB^*
=
0
.\eqbl
This is analogous to a force free equilibrium in an ordinary plasma \cite{Miller1981POF} where we have:
\bqbl
\lbq{eq:ForceFree}
\waJ\times\waB
=
0
.\eqbl
Substituting Eq.\,(\rfq{eq:defAStar}) and Eq.\,(\rfq{eq:defBStar}) into Eq.\,(\rfq{eq:RE_ForceBalance}) and compare with Eq.\,(\rfq{eq:ForceFree}), the only difference is the additional term $\waJ\times\left[\na\times\left(\R{p_\|}{q}\wsb\right)\right]$. We will see later on that this term will enter the right hand side of our new equilibrium equation.
In the axisymmetric system we considered here, the magnetic field $\waB$ and effective magnetic field $\waB^*$ could be expressed in the reduced form:
\bqbl
\lbq{eq:reducedB}
\waB
=
\na\gY\times\na\gj
+\baF\na\gj
,\eqbl
\bqbl
\lbq{eq:reducedBstar}
\waB^*
=
\na\gY^*\times\na\gj
+\baF^*\na\gj
.\eqbl
Here $\gY$ is the poloidal magnetic flux, the effective flux $\gY^*$ is related to the toroidal component of the effective vector potential $\waA^*$ which in turn is related to the canonical angular momentum of the runaway electrons. Meanwhile $\baF$ and $\baF^*$ are related to the toroidal field and the effective toroidal field respectively. The constant $\gY^*$ surfaces then defines the drift surfaces of the runaway electrons, which are the new characteristic surfaces analogous to the flux surfaces in an ordinary plasma equilibrium. Combining Eq.\,(\rfq{eq:waJBstar}) and Eq.\,(\rfq{eq:reducedBstar}), one could find
\bqbl
\lbq{eq:drift_surface_Bstar}
\waB^*\cdot\na\gY^*
=
0
,\eqbl
\bqbl
\lbq{eq:drift_surface_J}
\waJ\cdot\na\gY^*
=
0
.\eqbl
Hence the runaway electron trajectory and thus the current density flow along the drift surfaces. Note that this means $n_{RE}$ cannot be a flux function of $\gY^*$ in toroidal geometry due to the demand of current conservation, although $n_{RE}R$ is. This must be compensated by the cold bulk electrons which has much higher density, so that the quasi-neutrality could be maintained.

With Eq.\,(\rfq{eq:drift_surface_Bstar}) and Eq.\,(\rfq{eq:drift_surface_J}), it is evident that the runaway current dominant equilibrium can be defined by the drift surfaces, just like the flux surfaces define the ordinary plasma equilibrium. We then proceed to solve $\gY^*$.
In our toroidal symmetric system, realizing that 
\bqbl
\gm_0\waJ
=
\na\times\waB
=
-\gD^*\gY\na\gj
+\na\baF\times\na\gj
.\eqbl
Here $\gD^*\equiv R\R{\pa}{\pa R}\left(\R{1}{R}\R{\pa}{\pa R}\right)+\R{\pa^2}{\pa Z^2}$. 
So that
\bqbl
\gm_0\waJ\times\waB^*
=
-\R{\na\baF\baF^*}{R^2}
-\gD^*\gY^*\R{1}{R^2}\na\gY^*
=
0
,\eqbl
Hence we get a new (cold) Grad-Shafranov-like equation for the runway electron current:
\bqbl
\lbq{eq:REShafranov0}
\gD^\ast\gY
=
-\R{d\baF}{d\gY^*}\baF^*
.\eqbl
Note that here we have used the fact that the toroidal magnetic field $\baF$ is a function of the effective flux $\gY^*$. Unfortunately, this equation could not be directly solved since the RHS is not a function of $\gY$ alone, but are related to $\gY^*$. To solve it, one could try to convert this equation into a second order differential equation for $\gY^*$ instead, and make sure that the RHS only contains the function of $\gY^*$ and some spatial coordinates such as $R$.

To do this, we realize
\bqbl
\lbq{eq:REShafranov1}
\gD^\ast\gY
=
-\mu_{0}R^2\waJ\cdot\na\gj
=
-\gm_0R^2n_{RE}q\R{p_\|}{\gg m}\R{\waB^*}{B_\|^*}\cdot\na\gj
=
-\gm_0n_{RE}q\R{p_\|}{\gg m}\R{\baF^*}{B_\|^*}
.\eqbl
We note that Eq.\,(\rfq{eq:REShafranov0}) and Eq.\,(\rfq{eq:REShafranov1}) should be equivalent so that
\bqbl
\R{d\baF}{d\gY^*}
=
\gm_0n_{RE}q\R{p_\|}{\gg m}\R{1}{B_\|^*}
,\eqbl
which is a function of $\gY^*$. However, $\waF^{\ast}$ is not necessarily a function of $\gY^*$, and it is desirable to rewrite $\waF^{\ast}$ to separate the part which is a function of $\gY^*$ and the part which is not.

To do this, we rewrite Eq.\,(\rfq{eq:reducedBstar}) as
\bqbl
\lbq{eq:BstarSeparation}
\waB^*
&
=
&
\na\times \waA^*
=
\na\times \left(\waA+\R{p_\|}{q}\wsb\right)
\nonumber
\\
&
=
&
\left(\na\gY+\na\gz\right)\times\na\gj
+\left(\baF+\baG\right)\na\gj
.\eqbl
Here $\gz$ and $\baG$ represent the additional contribution from the mechanical momentum.
From Eq.\,(\rfq{eq:reducedB}) and Eq.\,(\rfq{eq:BstarSeparation}) we must have
\bqbl
\baG
=
R^2\left(\na\times\R{p_\|}{q}\wsb\right)\cdot\na\gj
.\eqbl
Realizing that under our current assumptions we have Eq.\,(\rfq{eq:p_paradot2}) so that $p_\|$ is a constant on a given drift surface. In this paper, as a first step, we are satisfied with consideration of the simplest case that $p_\|$ is also a constant over the whole volume.
We can then further simplify and obtain:
\bqbl
\baG
&
=
&
R^2\R{p_\|}{q}
\left[
\R{\na\times\waB}{B}\cdot\na\gj
-\R{\na B\times\waB}{B^2}\cdot\na\gj
\right]
\nonumber
\\
&
=
&
R^2\R{p_\|}{q}
\left[
-\R{\gD^*\gY}{BR^2}
-\R{\na B\times\waB}{B^2}\cdot\na\gj
\right]
.\eqbl
To the leading order, we have $\na B\simeq -\R{1}{R}B\na R$, so that we have
\bqbl
-\R{\na B\times\waB}{B^2}\cdot\na\gj
=
-\R{B}{R}\R{1}{B^2}\R{B_Z}{R}
=
-\R{B_Z}{BR^2}
.\eqbl
Since $B_Z=\R{1}{R}\R{\pa \gY}{\pa R}$, we have
\bqbl
\baG
&
=
&
\R{p_\|}{qB}
\left[
-R\R{\pa}{\pa R}\left(\R{1}{R}\R{\pa \gY}{\pa R}\right)
-\R{1}{R}\R{\pa \gY}{\pa R}
-\R{\pa^2\gY}{\pa Z^2}
\right]
\nonumber
\\
&
=
&
-\R{p_\|}{qB}
\left[
R\R{\pa B_Z}{\pa R}
+B_Z
+\R{\pa^2\gY}{\pa Z^2}
\right]
.\eqbl
It might be tempting to directly write $\baG=-\R{p_\|}{qB}\left[\R{\pa^2\gY}{\pa R^2}+\R{\pa^2\gY}{\pa Z^2}\right]$ here, but let's first compare the strength of $R\R{\pa B_Z}{\pa R}$ and $B_Z$. We notice that the variation of $B_Z$ is dominated by the contribution from the runaway current, and $B_Z$ will change sign over a length scale on the order of the minor radius $a$ as we go from the High Field Side (HFS) to the Low Field Side (LFS), should close flux surfaces still exist. Thus an order-of-magnitude analysis yields $R\R{\pa B_Z}{\pa R}\sim \R{R}{a}B_Z$ which is $\mathcal{O}\left(\ge^{-1}\right)$ compared with $B_z$. Keeping the leading order, we thus obtain
\bqbl
\baG
\simeq
-\R{p_\|}{qB}\gD^*\gY
.\eqbl
Hence we can express $\baF^*$ as
\bqbl
\lbq{eq:baF_baFStar}
\baF^*
=
\baF\left(\gY^*\right)
-\R{p_\|}{qB}\gD^*\gY
.\eqbl
We see that $\baF^*$ indeed is not a function of $\gY^*$ alone, but could be linked to $\baF$ and a function of the magnetic flux.

We could obtain the expression of $\nabla\gz$ in similar manner from Eq.\,(\rfq{eq:BstarSeparation}):
\bqbl
\na\gz
&
=
&
-R^2\left(\na\times\R{p_\|}{q}\wsb\right)\times\na\gj
\nonumber
\\
&
=
&
R^2\R{p_\|}{q}
\left[
\na\gj\times\R{\na\times\waB}{B}
-\na\gj\times\R{\na B\times\waB}{B^2}
\right]
\nonumber
\\
&
=
&
R^2\R{p_\|}{q}
\left[
\R{\na\baF}{BR^2}
+\R{B}{R}\R{\baF}{B^2}\R{\na R}{R^2}
\right]
\nonumber
\\
&
=
&
\R{p_\|}{qB}
\left[
\na\baF
+\R{\baF}{R}\na R
\right]
\nonumber
\\
&
\simeq
&
\R{p_\|}{q}\na R
.\eqbl
Here we have used the approximation $\frac{F}{B}=\frac{RF}{\sqrt{(\nabla\Psi)^{2}+F^{2}}}=R+\mathcal{O}\left(\ge^{2}\right)$ and the strong toroidal field assumption $R\na\baF\ll\baF$.
Finally, we can get the expression of $\nabla\Psi$
\bqbl
\lbq{eq:Psi_PsiStar}
\nabla\Psi = \nabla\Psi^\ast-\frac{p_{\parallel}}{q}\nabla{R} 
.\eqbl
Thus we recover the relationship between the effective flux $\gY^*$ and the magnetic flux $\gY$, which is consistent with Eq.\,(\rfq{eq:defAStar}):
\bqbl
\lbq{eq:Psi_PsiStar_2}
\Psi = \Psi^\ast -\frac{p_{\parallel}}{q}{R}
.\eqbl
Note here the contour of $\gY$ is our flux surface solution while the contour of $\gY^*$ is the drift surface solution, hence for finite $p_\|$ there will be a deviation between the flux surfaces and the drift surfaces.
Substituting Eq.\,(\rfq{eq:baF_baFStar}) and Eq.\,(\rfq{eq:Psi_PsiStar}) into Eq.\,(\rfq{eq:REShafranov1}). We immediately get the relationship between $\gD^*\gY^*$ and $\gD^*\gY$ according to Eq.\,(\rfq{eq:Psi_PsiStar}).
\bqbl
\gD^*\gY 
= 
\gD^*\gY^* + \R{p_\|}{qR}
.\eqbl
\bqbl
\gD^*\gY^*
+\R{p_\|}{qR}
=
-\gm_0n_{RE}q\R{p_\|}{\gg m}\R{1}{B_\|^*}\left[
\baF
-\R{p_\|}{qB}\left(\gD^*\gY^*+\R{p_\|}{qR}\right)
\right]
.\eqbl
This then yields
\bqbl
\left(
1-\gm_0n_{RE}\R{p_\|^2}{\gg m}\R{1}{B_\|^*B}
\right)\left(\gD^*\gY^*+\R{p_\|}{qR}\right)
=
-\gm_0n_{RE}q\R{p_\|}{\gg m}\R{\baF}{B_\|^*}
.\eqbl
We devide both sides by $-\frac{\gm_0n_{RE}p_\|}{\gg m B_\|^*}$ and obtain:
\bqbl
\left(
-\frac{\gg m B_\|^*}{n_{RE}\gm_0p_\|}+\frac{p_\|}{B}
\right)\left(\gD^*\gY^*+\R{p_\|}{qR}\right)
=
q\baF 
.\eqbl
Using the following approximation
\bqbl
B_\|^*B
=
\waB^*\cdot\waB
=
B^2
+\R{p_\|}{q}\gm_0\waJ\cdot\wsb
=
B^2
+\gm_0n_{RE}\R{p_\|^2}{\gg m}
,\eqbl
we could rewrite:
\bqbl
-\frac{\gg m B_\|^*}{n_{RE}\gm_0p_\|} +\frac{p_\|}{B}
= 
-\frac{B\gg m}{n_{RE}\gm_0p_\|}
.\eqbl
So that we finally have:
\bqbl
\lbq{eq:FinalREGSEq}
\gD^*\gY^*
=
-\gm_0n_{RE}qR\R{p_\|}{\gg m}
-\R{p_\|}{qR}
.\eqbl
Compared with ordinary Grad-Shafranov equation and taking into account of Eq.\,(\rfq{eq:Psi_PsiStar_2}), it can be seen that both our drift surface solution $\gY^*$ and flux surface solution $\gY$ would differ from the flux surface solution of the ordinary Grad-Shafranov equation for finite $p_\|$, while both of the first two solutions converge to the last in the limit of $p_\|\rightarrow 0$.
With given current density profile, $p_\|$ and boundary condition, the RHS of Eq.\,(\rfq{eq:FinalREGSEq}) is known and we could proceed to solve the equilibrium.

\section{The numerical solution of the runaway electron current dominated equilibrium}
\label{s:NumericalEq}
In this section, we present the numerical investigation of the runaway electron current under different control variables such as the parallel momentum, the current density profile and the externally applied vertical magnetic field. Specifically, in Section \ref{ss:ANVFS}, we introduced our numerical setup for the runaway electron current equilibrium solver. In Section \ref{ss:Dnaon}, the runaway current equilibrium with various parameters are calculated and plotted for both the ITER-like plasma and a smaller MAST-like plasma with complete up-down symmetric boundary condition, and found the change in the parallel momentum would result in horizontal current center displacement even with all the other parameters intact. In Section \ref{ss:Asymmetric boundary conditions}, we demonstrate that such horizontal displacement is accompanied by a vertical displacement if the boundary condition exhibit some up-down asymmetry.

\subsection{Numerical setup}
\label{ss:ANVFS}
Our numerical solver are based on the open source Grad-Shafranov solver for ordinary plasma equilibria Free-GS \cite{github}. The equilibrium equation is changed to that of our runaway current equilibrium in Section \ref{s:EquilibriumEq}, and the boundary condition is modified so that we have an ideally conducting boundary condition for the magnetic flux, plus whatever static externally applied poloidal magnetic field. 
Explicitly, the boundary condition for the magnetic flux in the up-down symmetric case is:
\bqbl
\gY_{b}
=
\R{1}{2}B_{z0}R^2
.\eqbl
This corresponds to the toroidal component of the magnetic vector potential $A_\gj=\R{1}{2}B_{z0}R$, which represents a constant vertical field $B_{z0}$. The corresponding boundary condition for $\gY^*$ is then
\bqbl
\gY_{b}^*
=
\R{1}{2}B_{z0}R^2
+\R{p_\|}{q}R
.\eqbl
This boundary condition represent the scenario that the current displacement within the vacuum vessel is faster than the resistive time of the wall, hence all magnetic flux changes will be screened, although the external field, which penetrate on a much longer timescale, is frozen in the wall. For simplicity, we only consider a rectangular boundary in this study. 
For the so-called ``MAST-like'' geometry, we have $R_{min}=0.1m$, $R_{max}=2.0m$, $Z_{min}=-2.0m$, $Z_{max}=2.0m$, while for the ``ITER-like'' geometry we have $R_{min}=4m$, $R_{max}=8.5m$, $Z_{min}=-4m$, $Z_{max}=5.0m$.
More detailed studies with realistic 2D first wall shape are left for future works. 

\begin{figure*}
\centering
\noindent
\btbl{c}
\parbox{5.0in}{
    \includegraphics[scale=0.4]{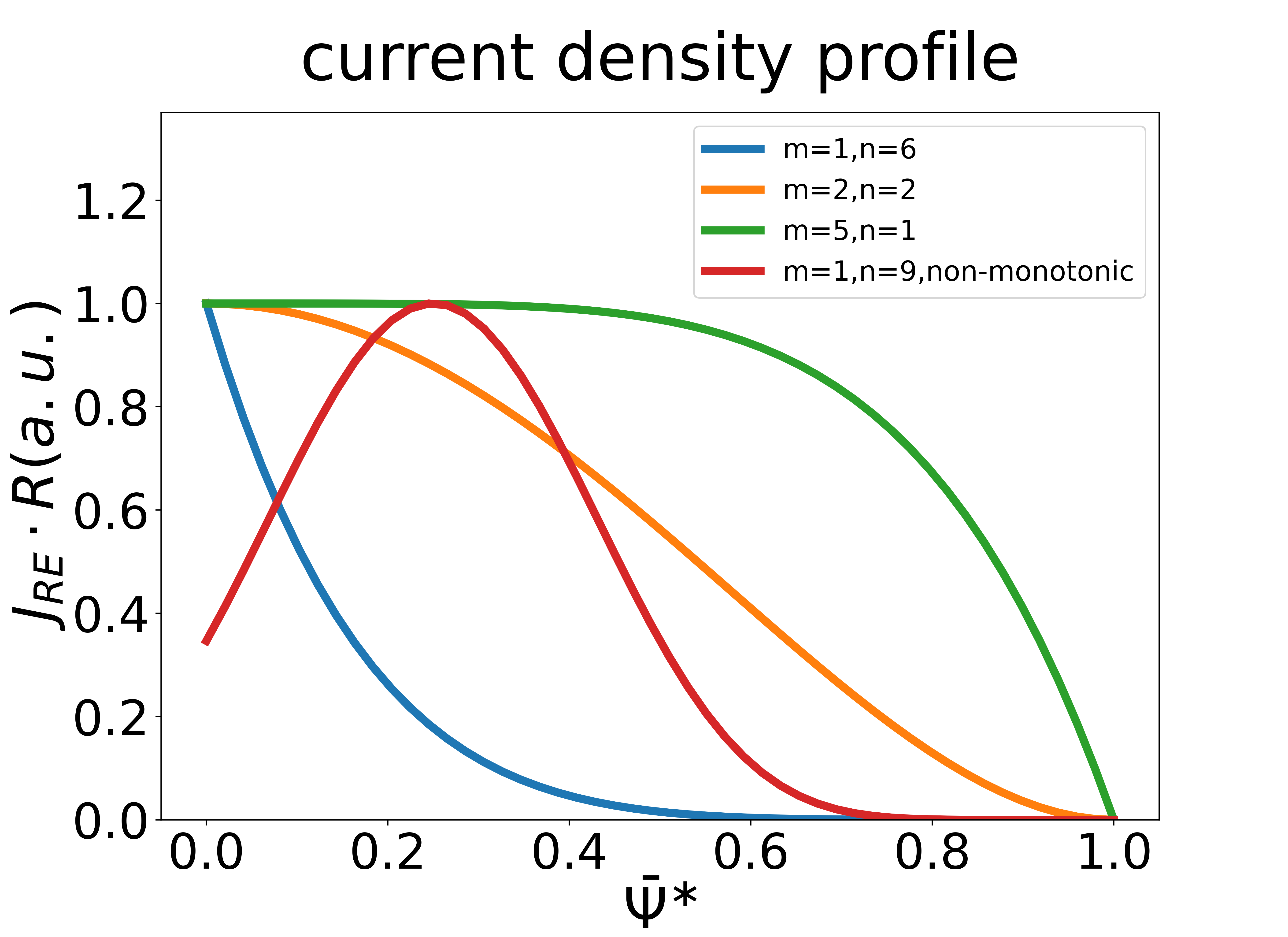}
}
\etbl
\caption{The shape of current density profiles according to Eq.\,(\rfq{eq:monotonicJ}) and Eq.\,(\rfq{eq:nonmonotonicJ}). The blue, red and orange curves represents three different monotonic current profiles, the green one is the non-monotonic one.}
\label{fig:nshape}
\end{figure*}

In this section, we mainly scan the parallel momentum $p_\|$ and a constant, uniform vertical external magnetic field $B_{z0}$ under different current density profiles.  
Such uniform vertical field of course is a simplification meant to provide a first estimation for our equilibrium solution. The realistic poloidal field or vertical field generated by the external coils are much more complicated in reality \cite{Gribov2015NF}, and we aim to include such realistic external coils in our future works.
We choose the monotonic current density profile to have the following shape:
\bqbl
J_\|R
=
n_{RE}q\R{p_\|}{\gg m}R
,\eqbl
\bqbl
\lbq{eq:monotonicJ}
n_{RE}R
=
n_{RE}\left(0\right)R_0\left(1-\bar{\gy^*}^m\right)^n
.\eqbl
Here $\bgy^*$ is the normalized effective flux. Several different parameters $m$ and $n$ are chosen to represent three kinds of distributions from the peaked profile to the flattened prole. Additionally, we consider one non-monotonic current profile to represent the possible hollowed runaway current spatial distribution\cite{Riemann2012POP,Zhou2013PPCF,Pusztai2022JPP}:
\bqbl
\lbq{eq:nonmonotonicJ}
n_{RE}R
=
n_{RE}\left(0\right)R_0\left[1-\bar{\gy^*}^2+\left(\bar{\gy^*}-\bar{\gy^*}^2\right)^m\right]^n
.\eqbl
All four kinds of current density profile are shown in Fig.\,\ref{fig:nshape} for comparison, with different colour representing different choice of $m$ and $n$ as well as the monotonic and non-monotonic profiles. Once the current profile shape and the total runaway current is chosen, the RHS of Eq.\,\rfq{eq:FinalREGSEq} is known.


Apart from the above constant, uniform $B_{z0}$ case, we also investigate an additional case with up-down asymmetry in the magnetic field boundary condition, to mimic the up-down asymmetry in the poloidal field coil in real tokamaks. In this simple study we only consider two poloidal field coils located at $R_u=5m$, $Z_u=7.5m$ and $R_d=6m$, $Z_d=-7.5m$ respectively, in addition to the uniform vertical field $B_{z0}$.
In this up-down asymmetric case, the boundary condition becomes:
\bqbl
\gY_{b}
=
\R{1}{2}B_{z0}R^2
+RA_{coil}
.\eqbl
Here $A_{coil}=\R{1}{2}\gm_0\sum_i{I_i\sum_{l=1,3,5,\cdots}{\R{P_l^1(0)}{l\left(l+1\right)}\left(\R{a_i}{r_i}\right)^{l+1}P_l^1\left(\cos{(\gq_i)}\right)}}$ is the total toroidal magnetic vector potential contribution from the external PF coils, where $a_i$ is the radius of coil $i$, $r_i$ is the distance from the boundary point to the coil centre. Here we assume $r_i>a_i$. Further, $\gq_i$ is the poloidal angle of the boundary point related to the coil centre, $I_i$ is the coil current and $P_l^1$ is the first order associated Legendre polynomial.
The corresponding boundary condition for $\gY^*$ is then
\bqbl
\gY_{b}^*
=
\R{1}{2}B_{z0}R^2
+RA_{coil}
+\R{p_\|}{q}R
.\eqbl
With those boundary conditions, we can proceed to solve the runaway electron current dominant equilibrium.

\subsection{Numerical results for up-down symmetric boundary condition}
\label{ss:Dnaon}

We demonstrate here the runaway current equilibria calculated using Eq.\,\rfq{eq:FinalREGSEq} and the up-down symmetric boundary conditions described in Section \ref{ss:ANVFS}. 
\begin{figure*}
\centering
\noindent
\btbl{cc}
\parbox{2.5in}{
    \includegraphics[scale=0.175]{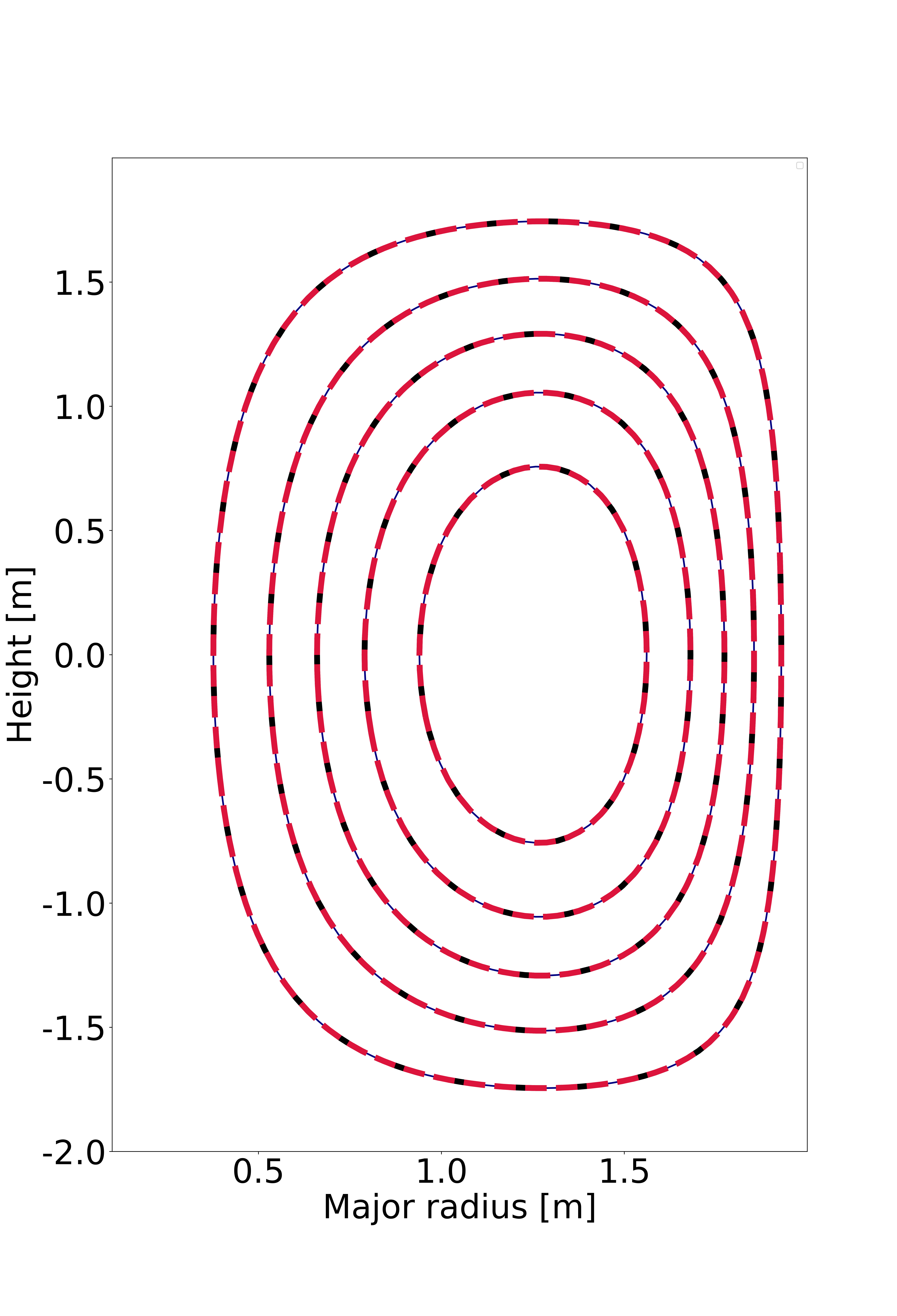}
}
&
\parbox{2.5in}{
	\includegraphics[scale=0.175]{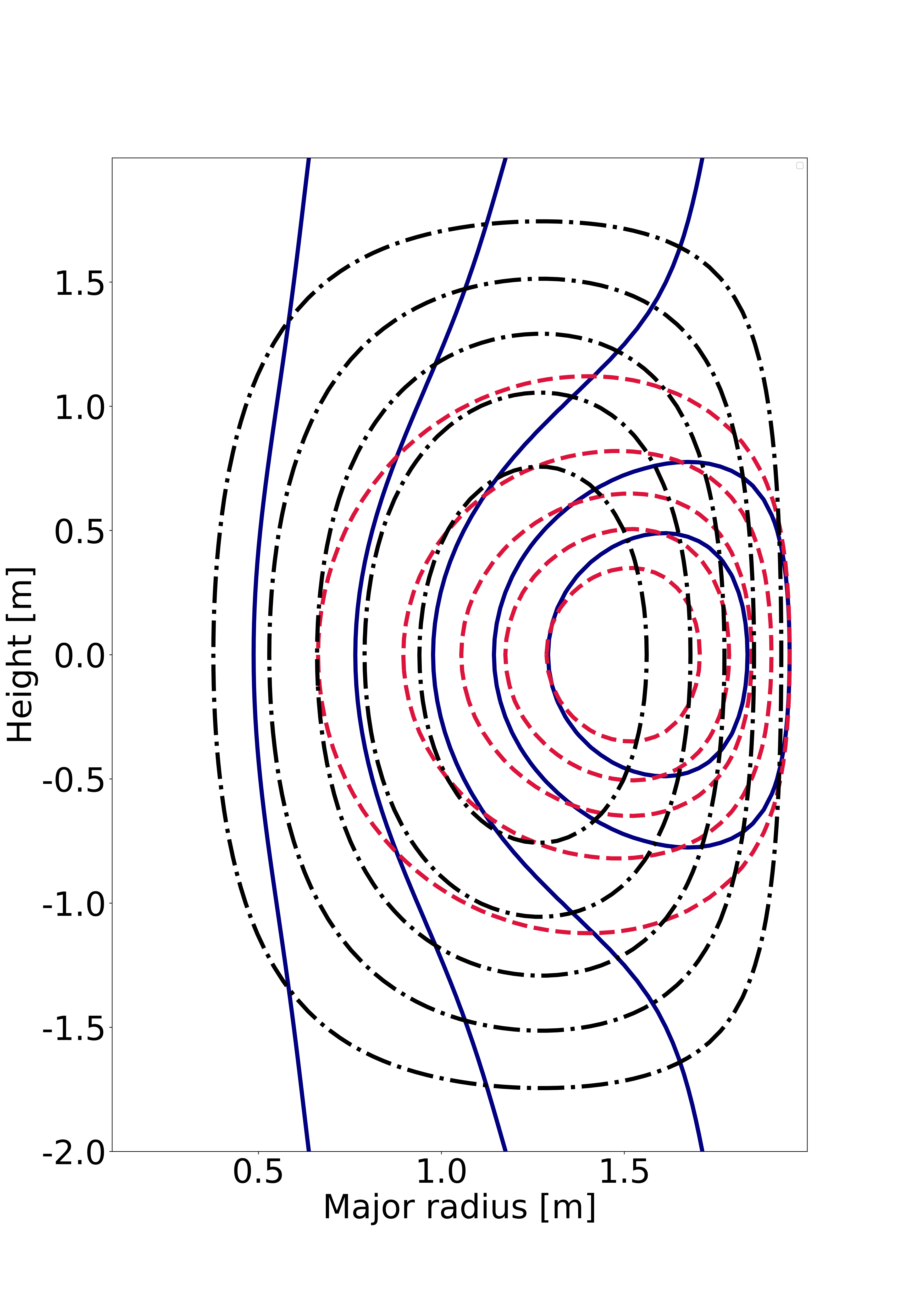}
}
\\
(a)&(b)
\etbl
\caption{The comparison between the flux surface (red dashed lines) and the drift surface (blue solid lines) of our new equilibrium solution as well as that of an ordinary Grad-Shafranov equation (black chained lines) with MAST-like geometry, $B_{z0}=0$ and $I_p=200kA$ for (a) $\gg=0$ and (b) $\gg=20$.}
\label{fig:benchmark}
\end{figure*}

To demonstrate the difference between our new equilibrium solution and the ordinary Grad-Shafranov solution, as well as their convergence in the limit of $p_\|\rightarrow 0$, we hereby show two benchmark cases with FreeGS results using the ideal boundary MAST-like geometry with $B_{z0}=0$ and $I_p=200kA$. The flux surface and the drift surface of our equilibrium solution as well as the flux surface solution from FreeGS are compared in Fig.\,\ref{fig:benchmark} for both the $p_\|=0$ and the $p_\|=20mc$ cases. The black chained lines represent the Grad-Shafranov solution, the dark blue solid lines represent the new drift surface solution and the red dashed lines represent the new flux surface solution. In Fig.\,\ref{fig:benchmark}(a) where we have $p_\|=0$, exact agreement is found between all three surfaces as the contours of the three solutions overlap with each other. This demonstrates we could recover the ordinary force free Grad-Shafranov solution in the low electron momentum limit. As the electron momentum increase, however, deviation between the solutions becomes gradually more pronounced, as can be seen in Fig.\,\ref{fig:benchmark}(b) where we have $p_\|=20mc$. Here the Grad-Shafranov solution is unchanged, as the parallel momentum of the current carrier does not enter the equilibrium equation explicitly. Meanwhile, both the flux surfaces and the drift surfaces of our new equilibrium solution exhibit an outward displacement as the electron parallel momentum increases. The deviation between the drift surface and the flux surface solution also begin to manifest itself, as this deviation is proportional to $p_\|$ as we discussed in Section \ref{s:EquilibriumEq}.

\begin{figure*}
\centering
\noindent
\btbl{cc}
\parbox{2.5in}{
    \includegraphics[scale=0.23]{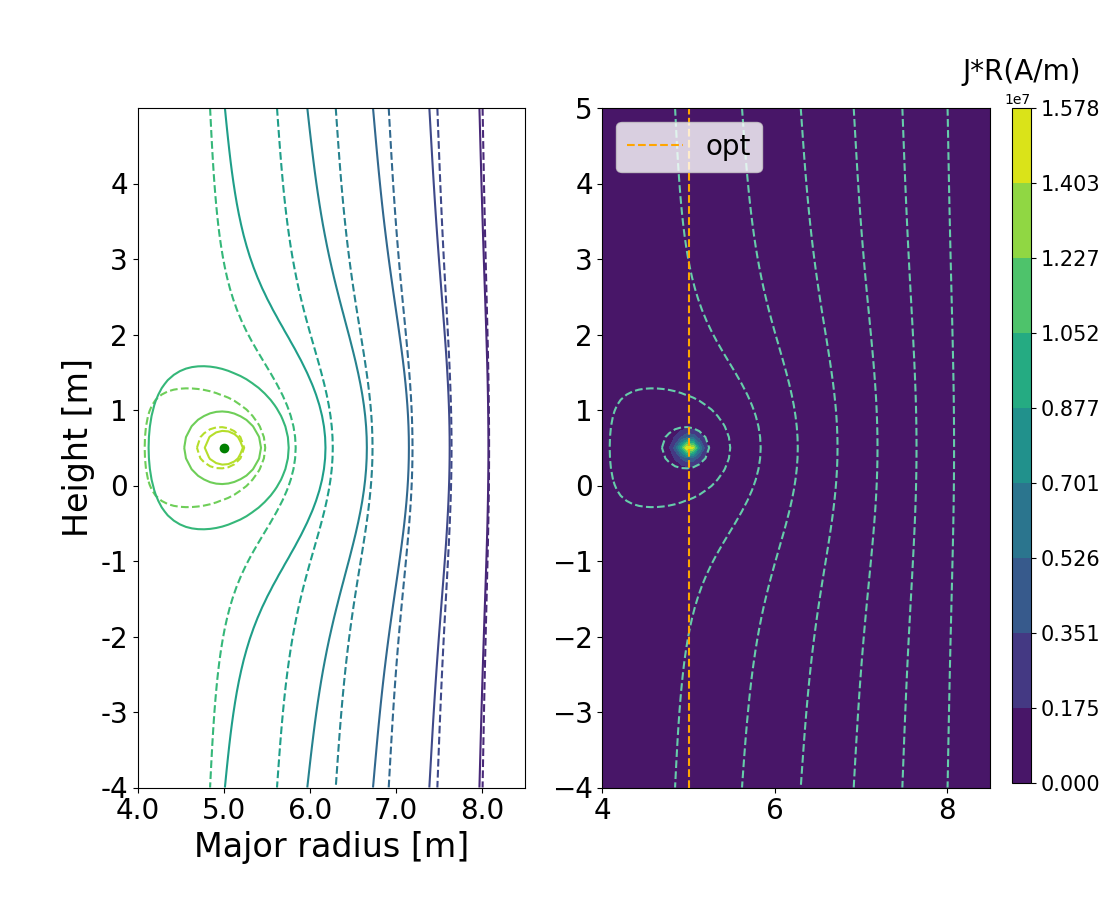}
}
&
\parbox{2.5in}{
	\includegraphics[scale=0.23]{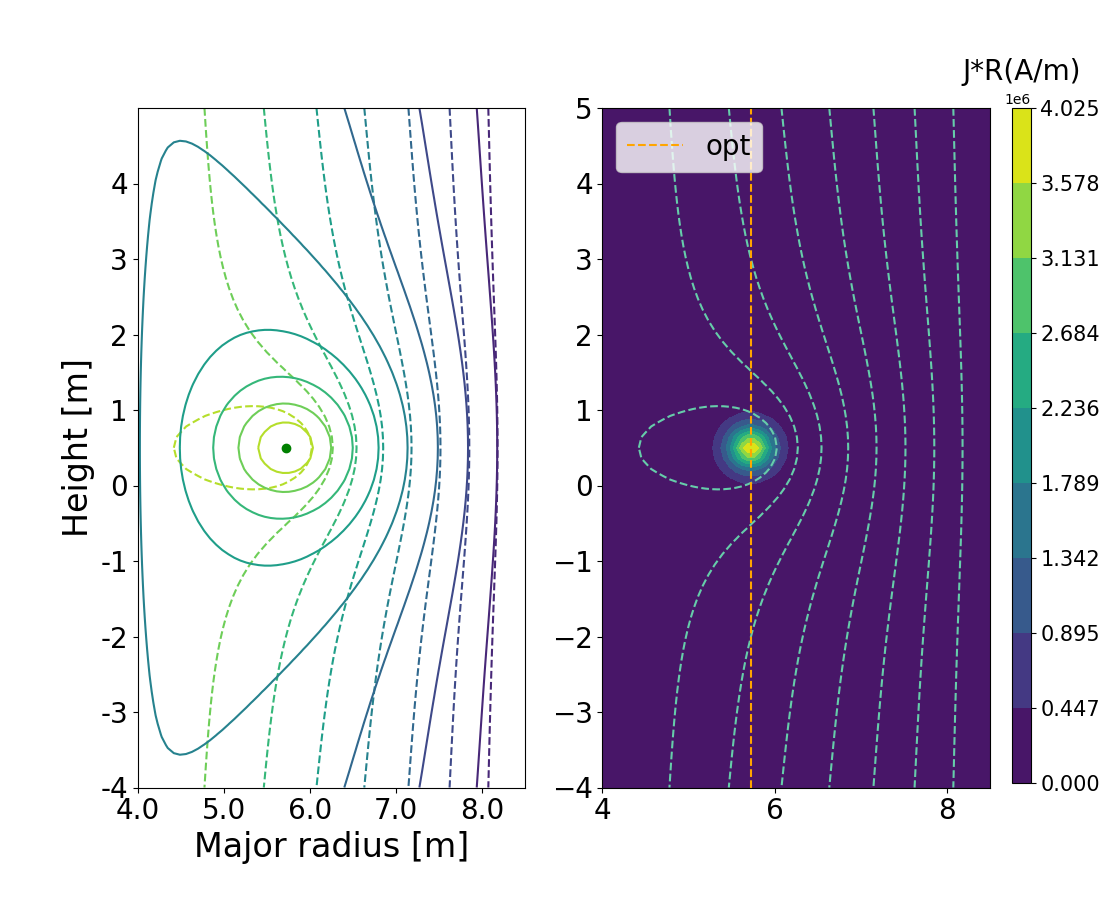}
}
\\
(a)&(b)
\\
\parbox{2.5in}{
    \includegraphics[scale=0.23]{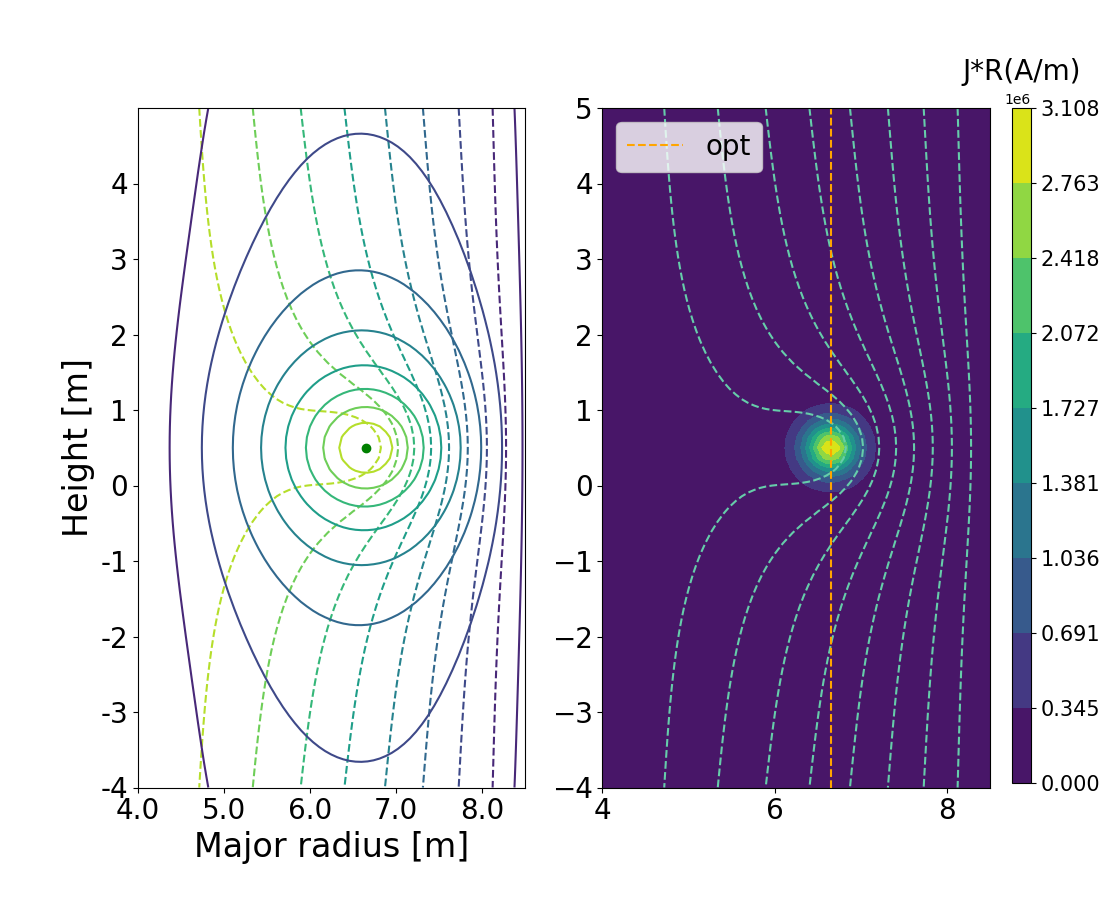}
}
&
\parbox{2.5in}{
    \includegraphics[scale=0.23]{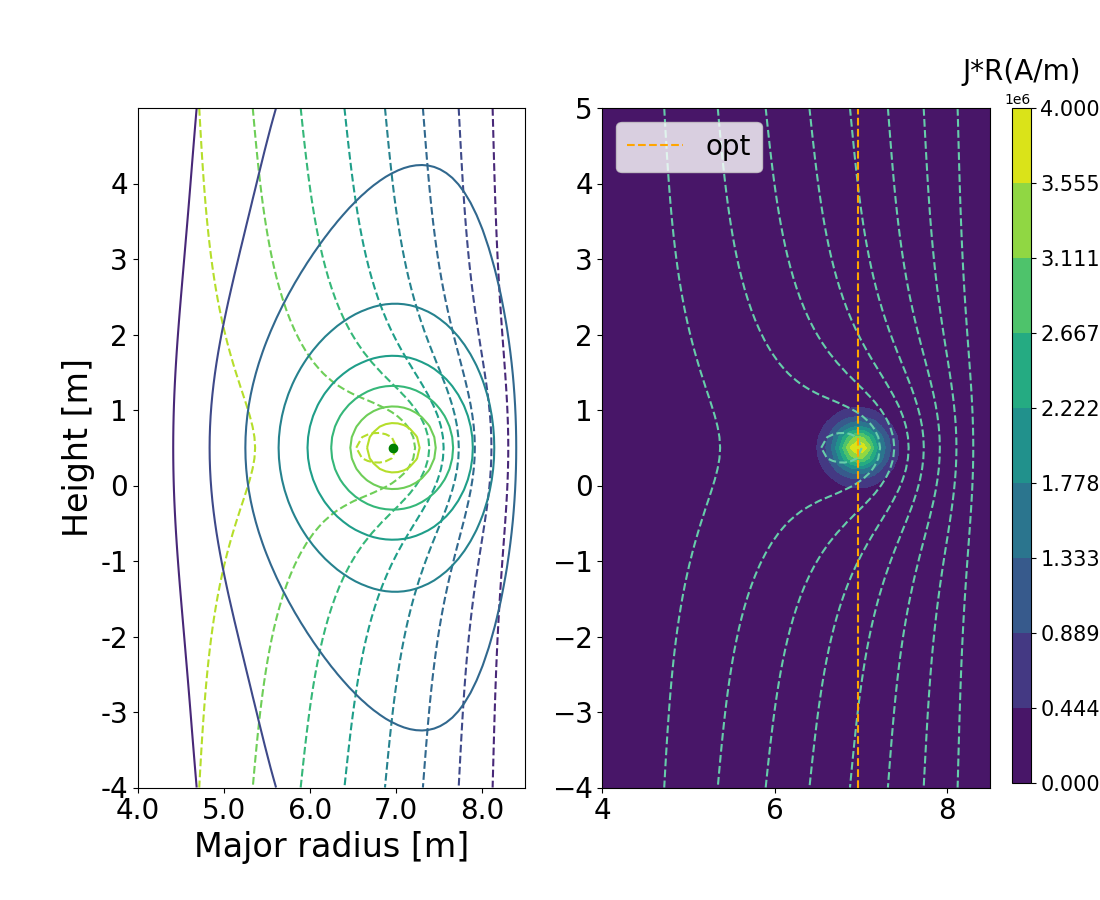}
}
\\
(c)&(d)
\etbl
\caption{The runaway current equilibrium with $B_{z0}=-0.045T$, and (a) $\gg=56$, (b) $\gg=116$, (c) $\gg=176$ and (d) $\gg=200$, respectively. The current profile is monotonic with $m=1$ and $n=6$ and total runaway electron current $I_{RE}=200kA$. The solid contours represent the drift surfaces and the dashed contours represent the flux surfaces.}
\label{fig:2e51_6}
\end{figure*}

We then look at the peaked current profile case described by Eq.\,\rfq{eq:monotonicJ} with $m=1$ and $n=6$ and total runaway electron current $I_{RE}=200kA$ but with varying $p_\|$. The equilibrium solution for several chosen $p_\|$ and $B_{z0}=-0.045T$ as is shown in Fig.\,\ref{fig:2e51_6}, where the solid contours represent the drift surfaces and the dashed contours represent the flux surfaces. 
This choice of $B_{z0}$ value is on the same order of magnitude as one would expect to maintain the horizontal position of the current centre for an ordinary plasma \cite{WessonBook}, although it is not exactly the same due to the contribution from the runaway electron parallel momentum and the back reaction from the ideal wall.
The orange vertical dashed line mark the horizontal position of the current center. The color bar for the right figure in each subplot represents the value of $J_\|R$ which is a function of $\gY^*$. Note that such a centrally peaked current profile is likely to trigger some kink instability for the runaway current, but this is not the concern of our 2D equilibrium study here and we leave detailed stability studies for future works. From Fig.\,\ref{fig:2e51_6}(a) to (d), it can be seen that even with exactly the same vertical field and the same runaway electron current, the current center shows apparent displacement as the parallel momentum changes, in agreement with previous simplified model \cite{Di2016POP}. Such parallel momentum change could be caused by many mechanism, such as the collision with background species \cite{Hesslow2017PRL}, radiative drag \cite{Martin-Solis2008POP}, kinetic instabilities \cite{Liu2018PRL} or toroidal electric field. In the small $p_\|$ limit, the deviation between the drift surface and the flux surface is small, while such deviation consistently grows as $p_\|$ increases. In the large $p_\|$ limit, although there is only very little closed flux surface remains, significant closed drift surfaces still exist, and the runaway electrons remain well confined even in the open field line region. This feature cannot be captured with the ordinary Grad-Shafranov equation since the current flows along the flux surfaces, and there can not be closed current surfaces in open field line region.
This runaway electron drift is closely associated with the aforementioned current center displacement, as the drifted runaway electrons are also the main current carrier, thus any runaway electron drift would also move the current channel, which in turn moves the magnetic flux, which causes further runaway electron drift. Thus we see inward current displacement for decreasing runaway electron momentum, and outward displacement for growing momentum.

\begin{figure*}
\centering
\noindent
\btbl{cc}
\parbox{2.5in}{
    \includegraphics[scale=0.225]{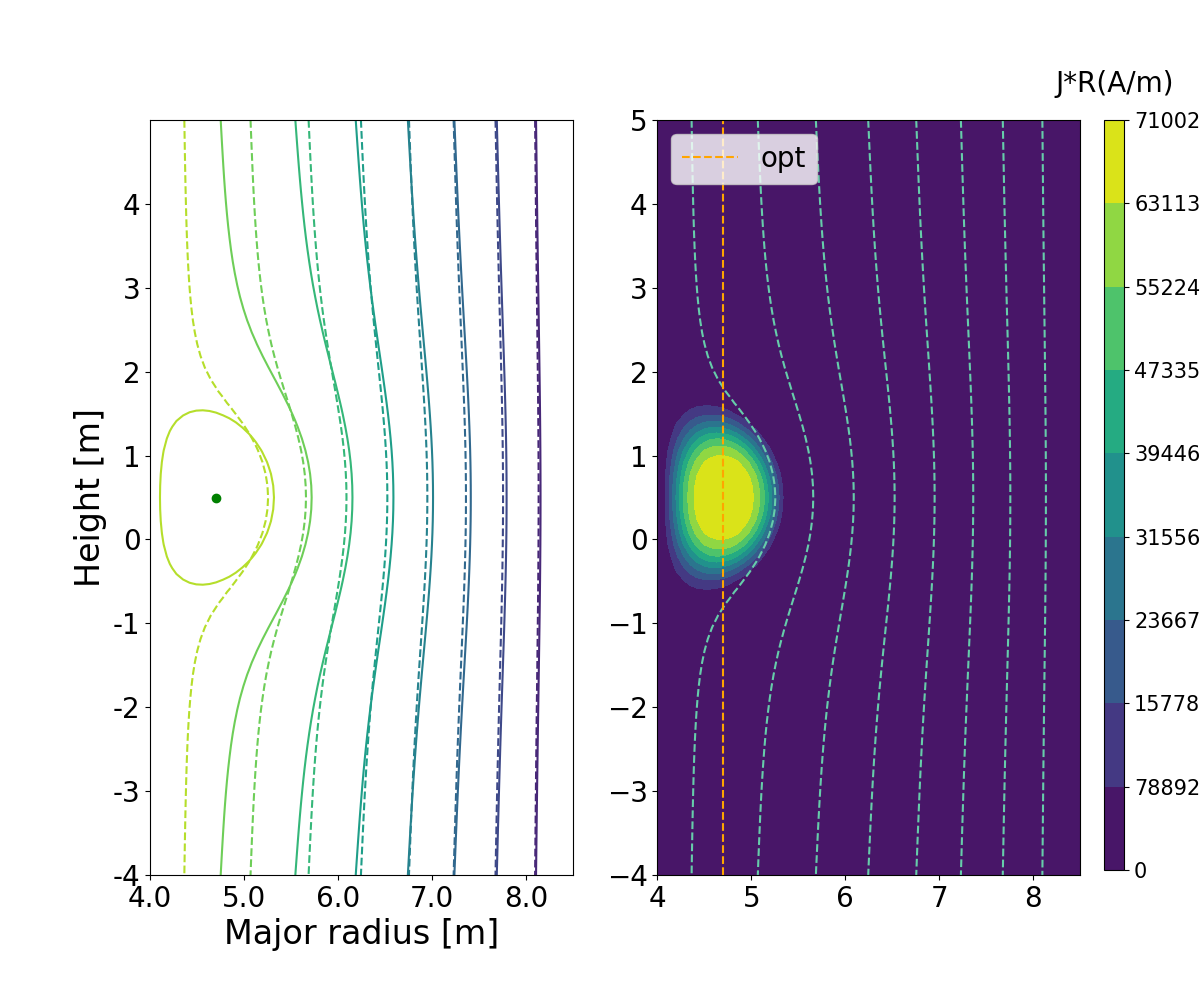}
}
&
\parbox{2.5in}{
	\includegraphics[scale=0.225]{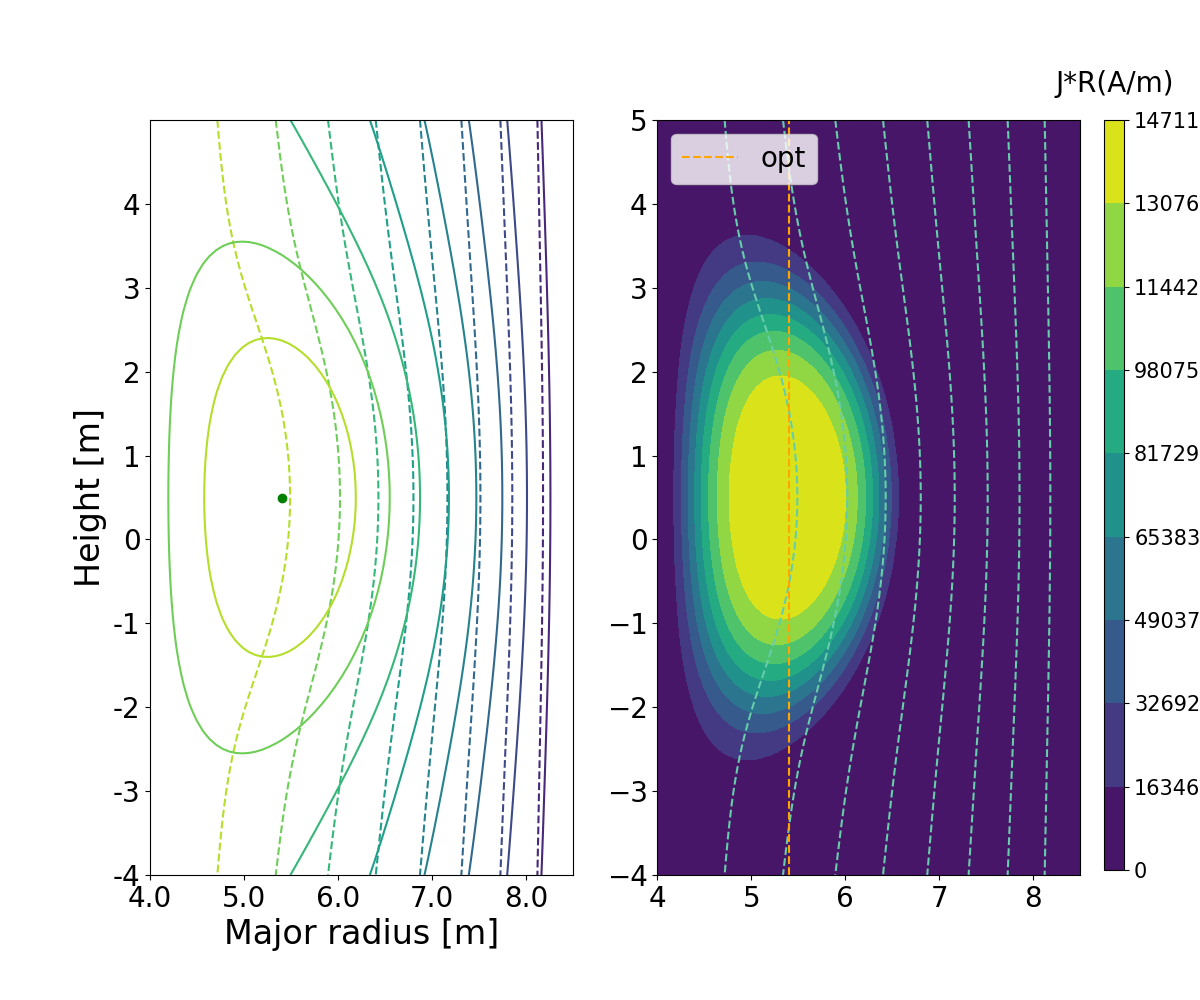}
}
\\
(a)&(b)
\\
\parbox{2.5in}{
    \includegraphics[scale=0.225]{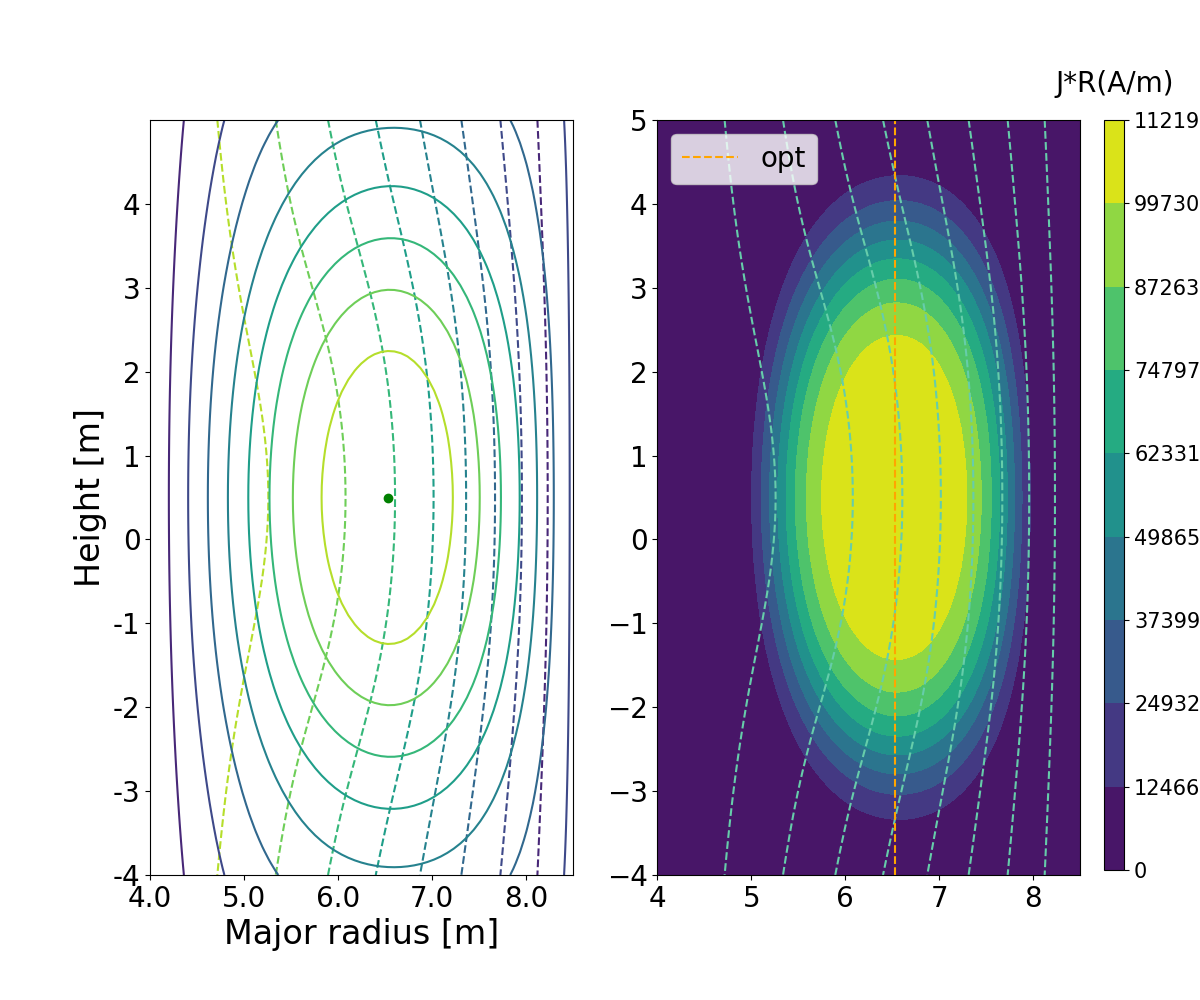}
}
&
\parbox{2.5in}{
    \includegraphics[scale=0.225]{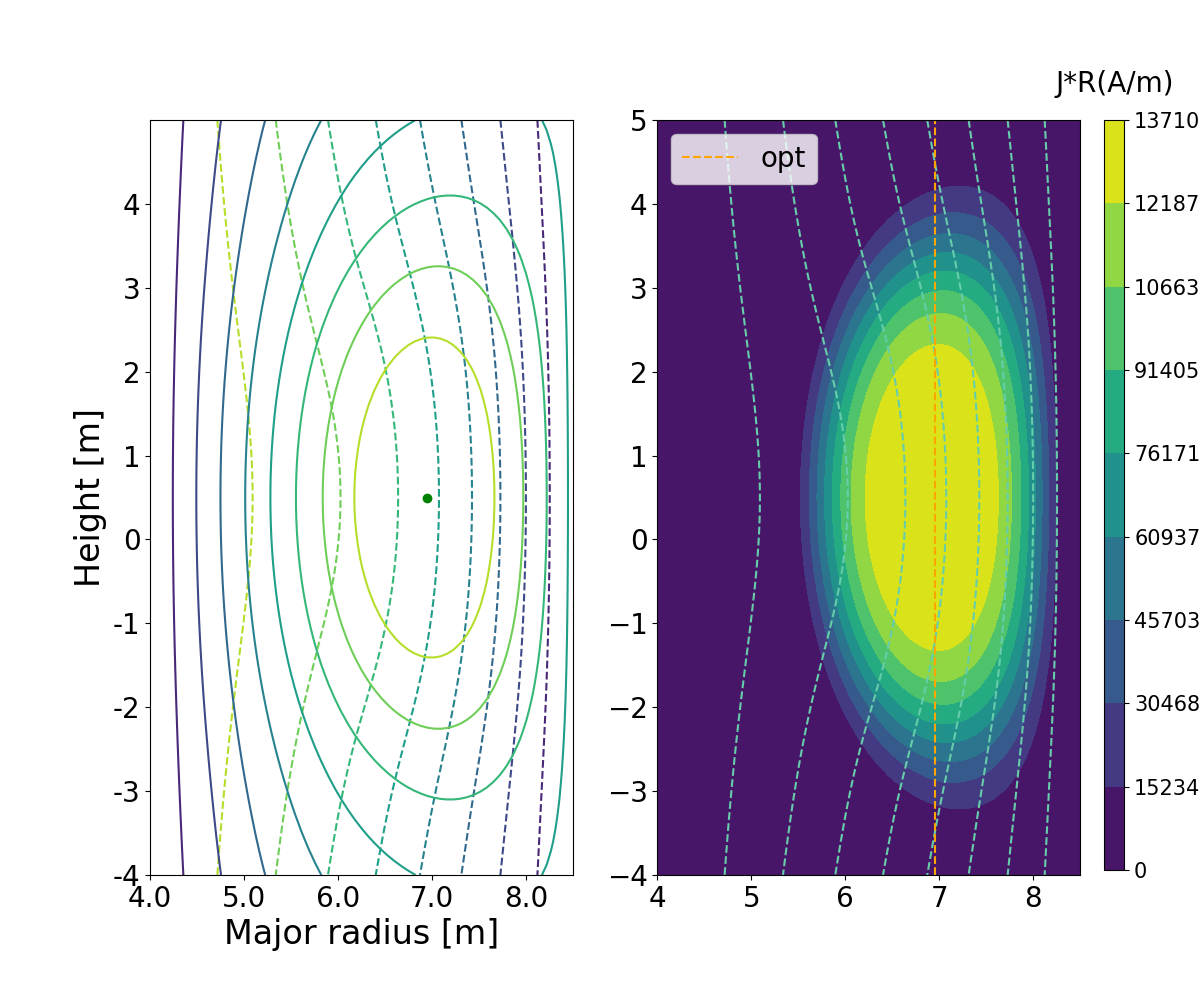}
}
\\
(c)&(d)
\etbl
\caption{The runaway current equilibrium with $B_{z0}=-0.045T$, and (a) $\gg=56$, (b) $\gg=116$, (c) $\gg=176$ and (d) $\gg=200$, respectively. The current profile is monotonic with $m=2$ and $n=2$ and total runaway electron current $I_{RE}=200kA$. The solid contours represent the drift surfaces and the dashed contours represent the flux surfaces. }
\label{fig:2e52_2}
\end{figure*}


As a comparison, we show another example of runaway current equilibrium with more flattened current profile $m=2$ and $n=2$, as is plotted in Fig.\,\ref{fig:2e52_2}. Similar horizontal current displacement caused by the change of the parallel momentum alone can be seen. Consistent with the $m=1$ and $n=6$ case, closed runaway electrons drift surfaces exist where field lines are open. Compared with Fig.\,\ref{fig:2e51_6}, however, Fig.\,\ref{fig:2e52_2} show significant shrinking of the current channel as the horizontal current displacement occurs. This is due to the shrinking of our ``last closed drift surface'' as the runaway electron current move towards the wall, which mimics the scraping-off of the edge runaway electron current and it redistribution to the core by the inductive electric field in real current displacement event. Comparing the value of $p_\|$ at similar current center position in Fig.\,\ref{fig:2e51_6} and Fig.\,\ref{fig:2e52_2}, the more flattened current profile requires less change in $p_\|$ for the same current displacement, suggesting that their current center position is more sensitive to the parallel momentum change. This may be due to the fact that the flattened current profile case experience more significant current shrinking compared with the peaked current profile case, which result in stronger change in their internal inductance.

\begin{figure*}
\centering
\noindent
\btbl{cc}
\parbox{2.5in}{
    \includegraphics[scale=0.225]{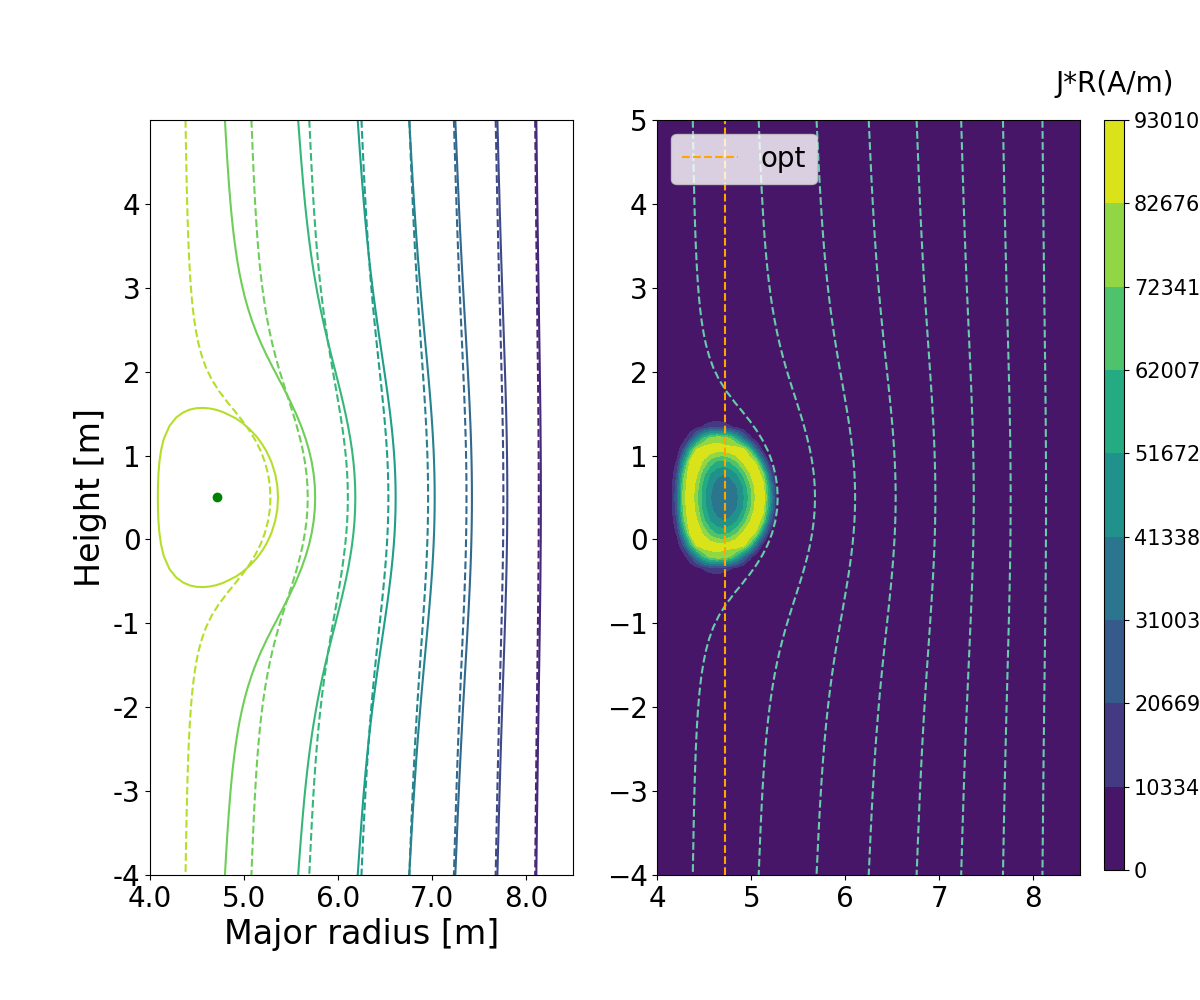}
}
&
\parbox{2.5in}{
	\includegraphics[scale=0.225]{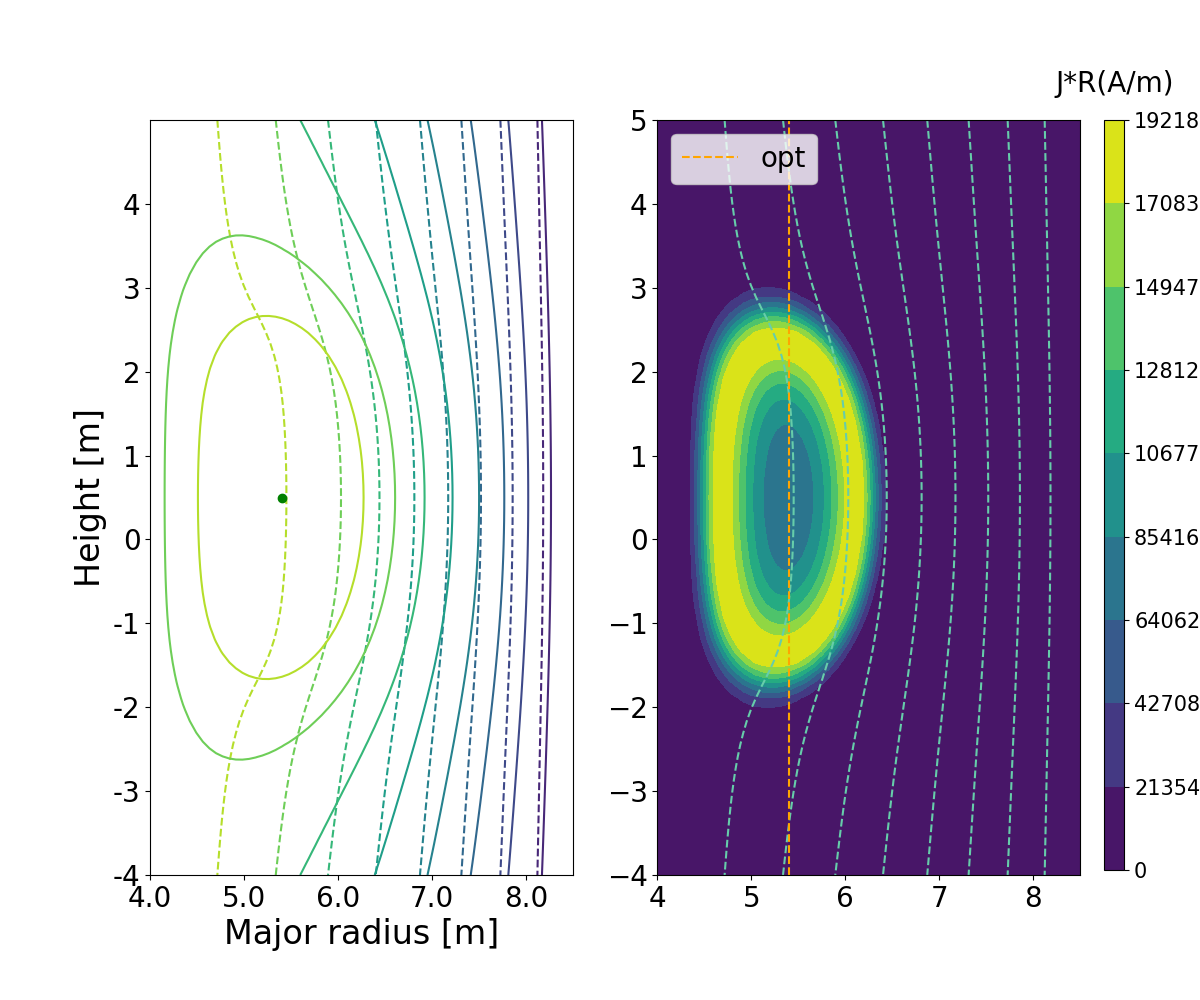}
}
\\
(a)&(b)
\\
\parbox{2.5in}{
    \includegraphics[scale=0.225]{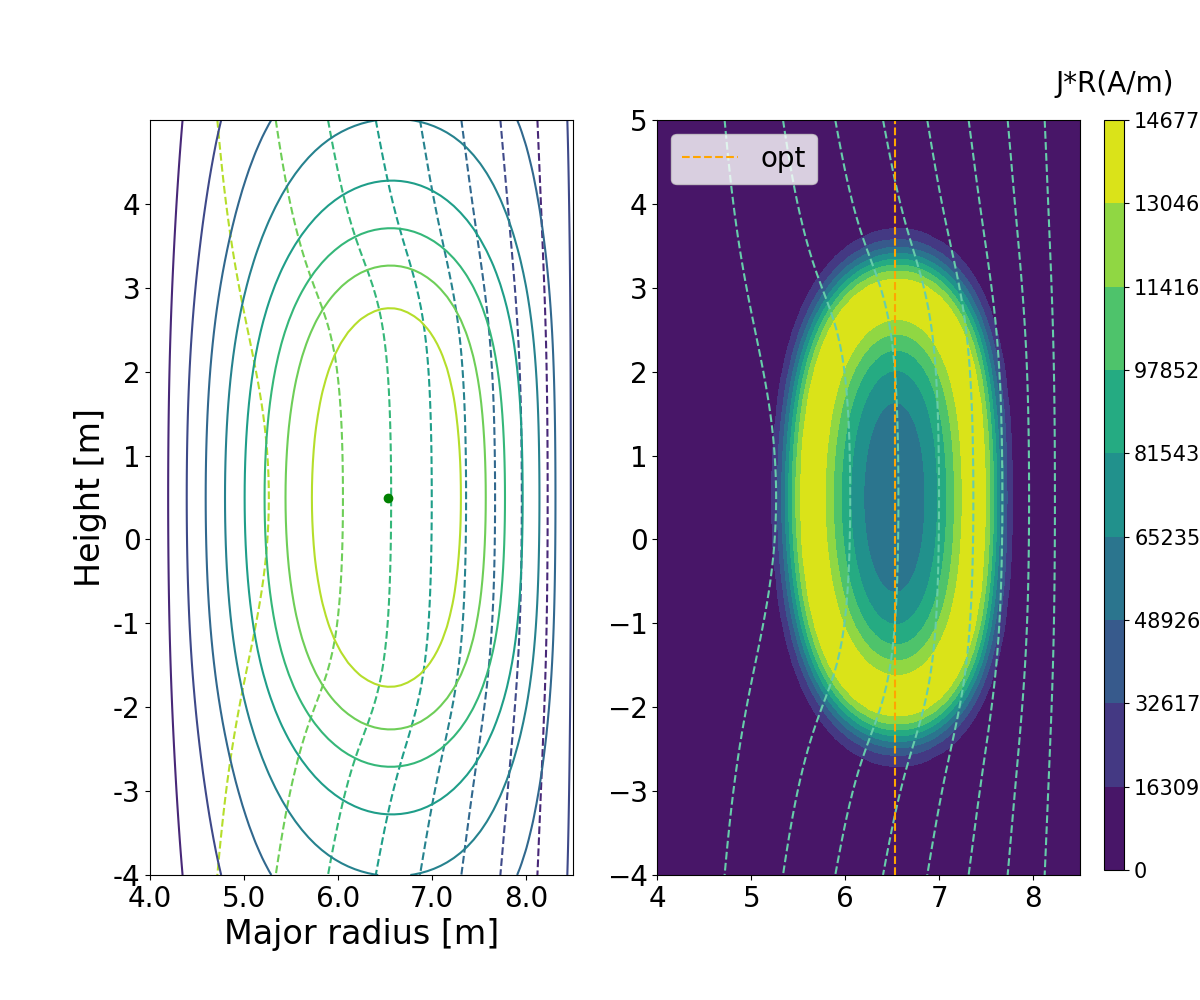}
}
&
\parbox{2.5in}{
	\includegraphics[scale=0.225]{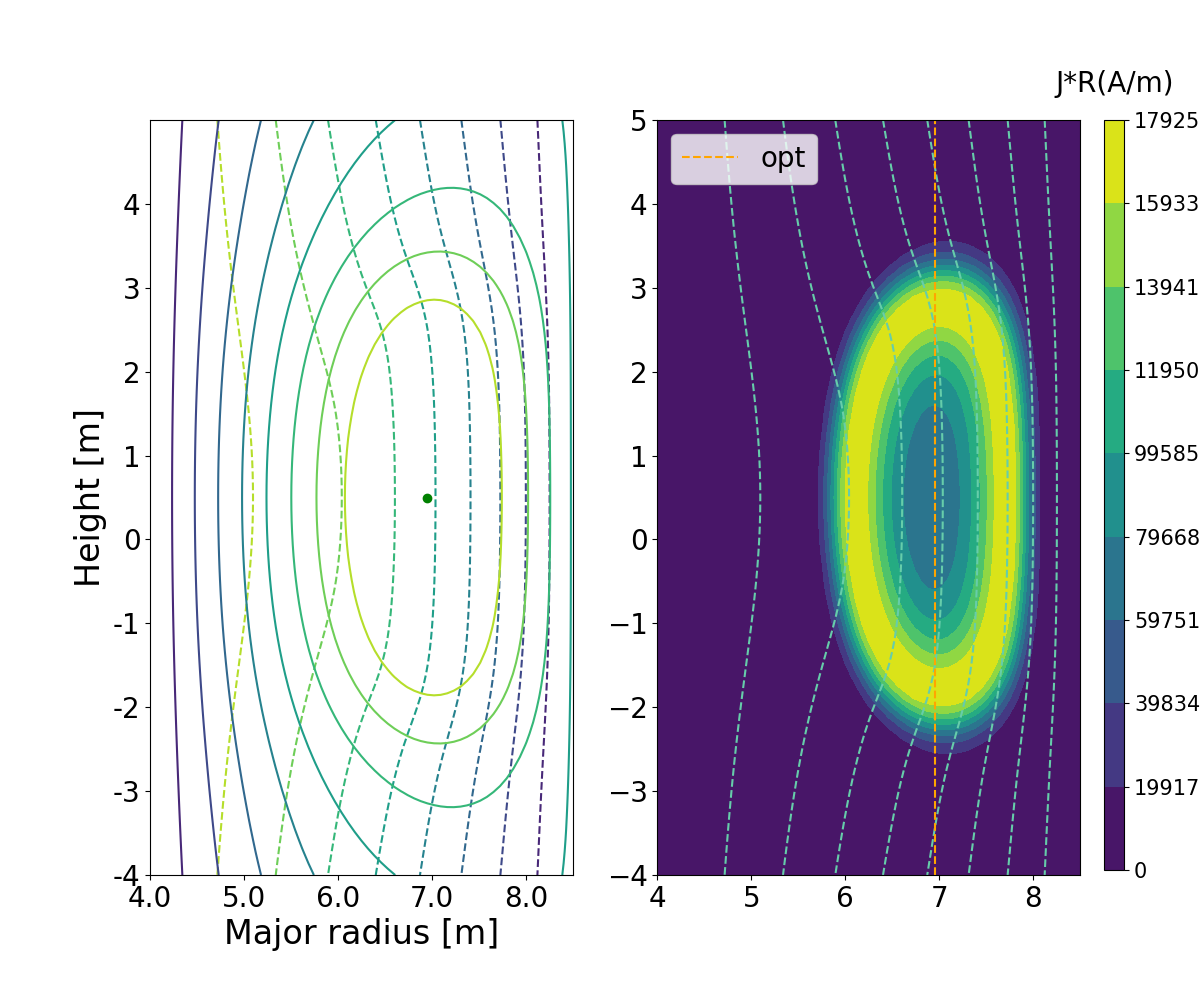}
}
\\
(c)&(d)
\etbl
\caption{The runaway current equilibrium with $B_{z0}=-0.045T$, and (a) $\gg=56$, (b) $\gg=116$, (c) $\gg=176$ and (d) $\gg=200$, respectively. The current profile is non-monotonic with $m=1$ and $n=9$ and total runaway electron current $I_{RE}=200kA$. The solid contours represent the drift surfaces and the dashed contours represent the flux surfaces. }
\label{fig:2e51_9}
\end{figure*}

We now take a look at the non-monotonic current profile case with $m=1$ and $n=9$ described by Eq.\,\rfq{eq:nonmonotonicJ}, the equilibrium solutions for several choices of $p_\|$ are shown in Fig.\,\ref{fig:2e51_9}. Comparing Fig.\,\ref{fig:2e51_9} and Fig.\,\ref{fig:2e52_2}, this hollowed current profile case shows similar displacement with the $m=2$, $n=2$ monotonic current profile shape for the same $p_\|$ change. A similar scraping-off of the edge current density could be seen comparing Fig.\,\ref{fig:2e51_9}(a) and (b) with Fig.\,\ref{fig:2e51_9}(c). For our case studied here, the non-monotonic feature does not seems to have significant contribution to the susceptibility of the current center displacement.

\begin{figure*}
\centering
\noindent
\btbl{cc}
\parbox{2.5in}{
    \includegraphics[scale=0.165]{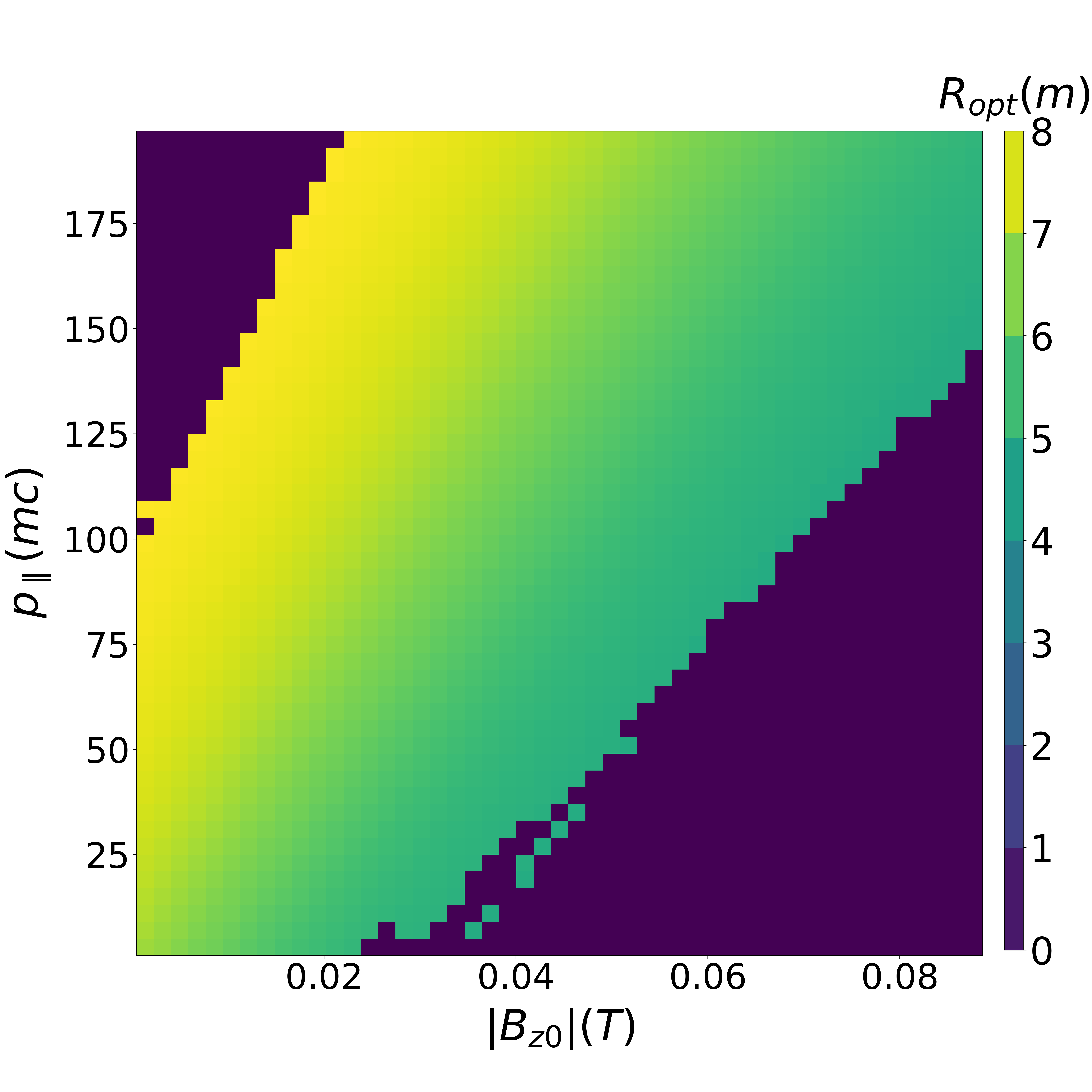}
}
&
\parbox{2.5in}{
	\includegraphics[scale=0.165]{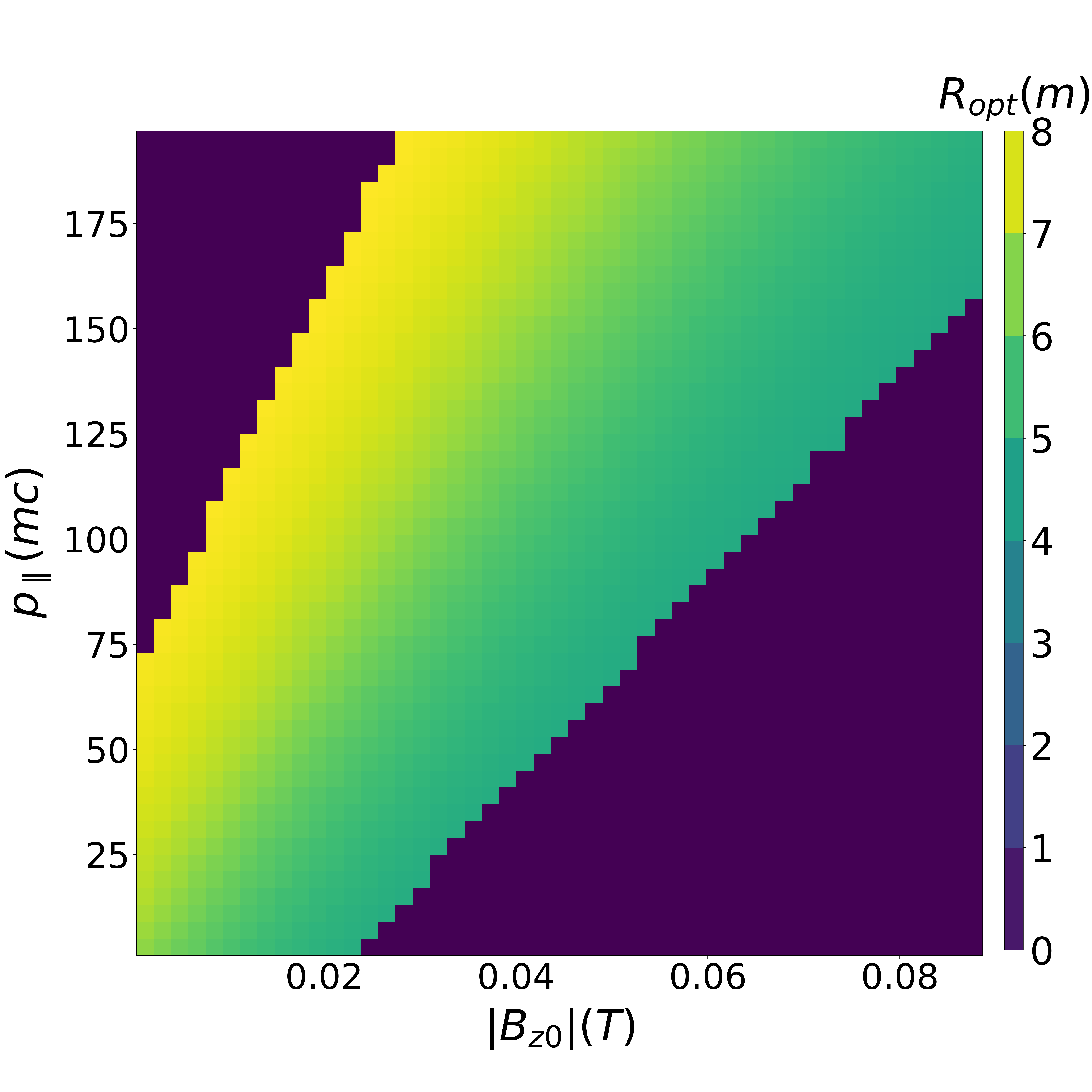}
}
\\
(a)&(b)
\\
\parbox{2.5in}{
    \includegraphics[scale=0.165]{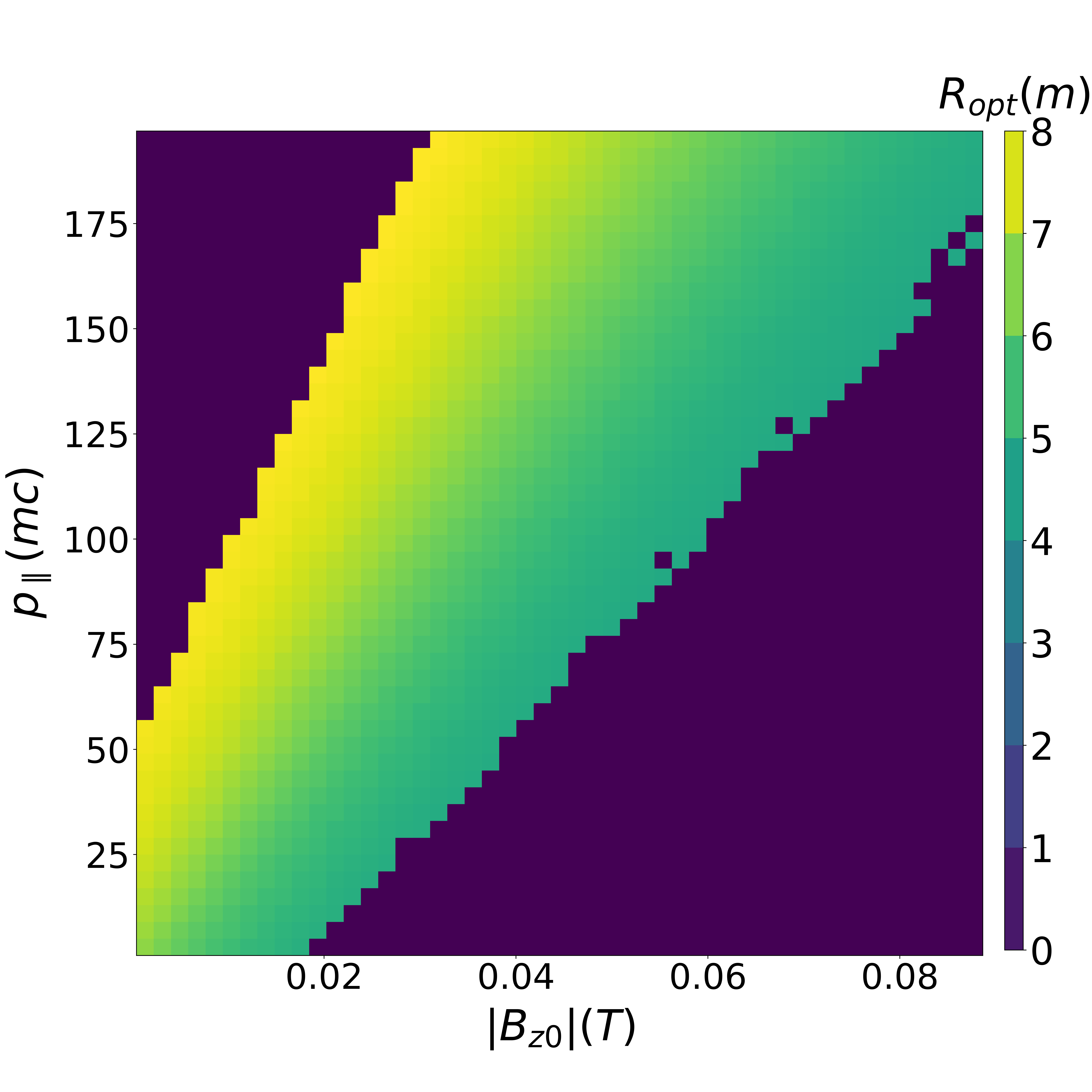}
}
&
\parbox{2.5in}{
	\includegraphics[scale=0.165]{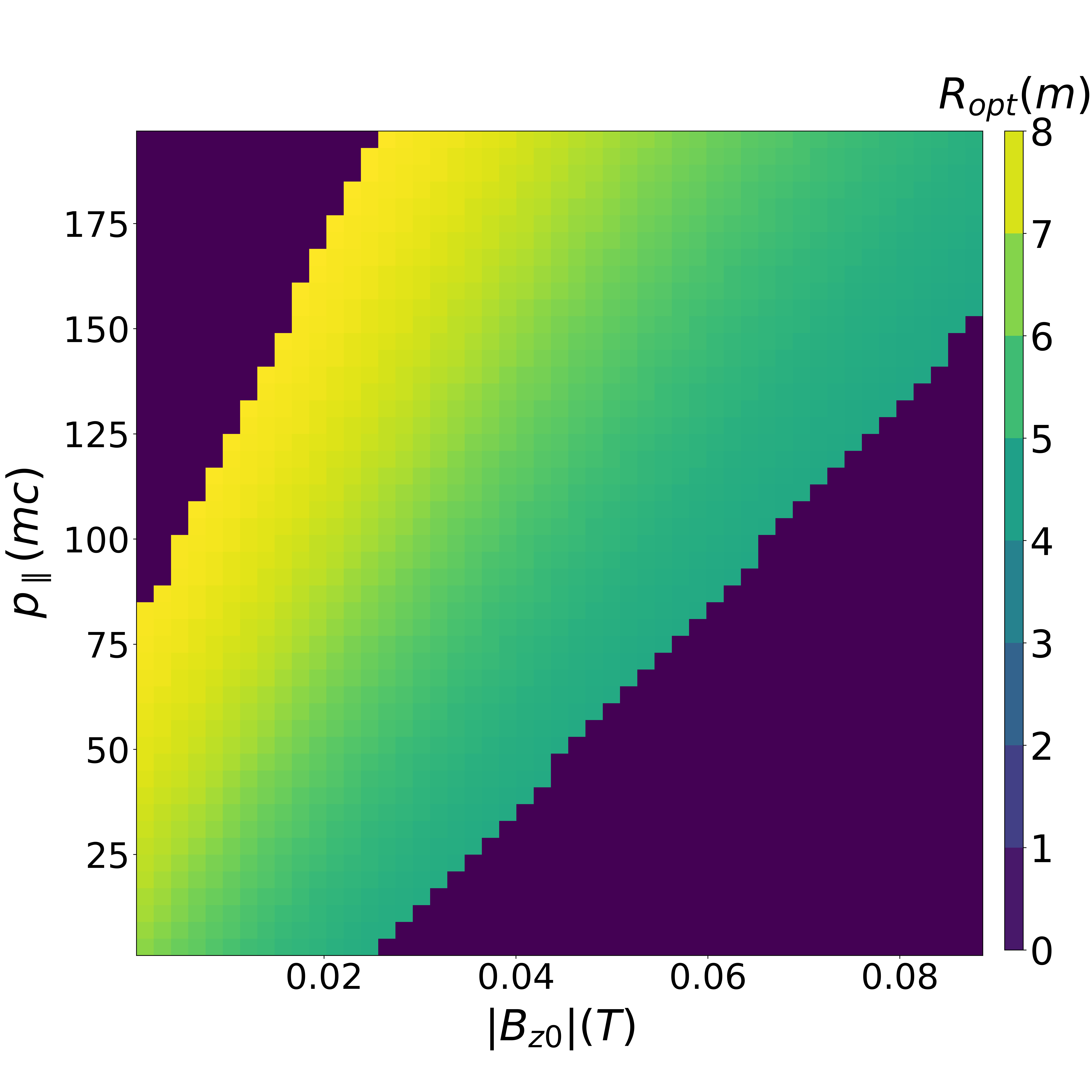}
}
\\
(c)&(d)
\etbl
\caption{The current center major radius position as a function of $p_\|$ and $B_{z0}$ for (a) the monotonic $m=1$, $n=6$ case, (b) the monotonic $m=2$, $n=2$ case, (c) the monotonic $m=5$, $n=1$ case and (d) the non-monotonic $m=1$, $n=9$ case. The dark blue colour represent where the equilibrium cannot be found due to too much current displacement.}
\label{fig:allPBO}
\end{figure*}

\begin{figure*}
\centering
\noindent
\btbl{c}
\parbox{5.0in}{
    \includegraphics[scale=0.55]{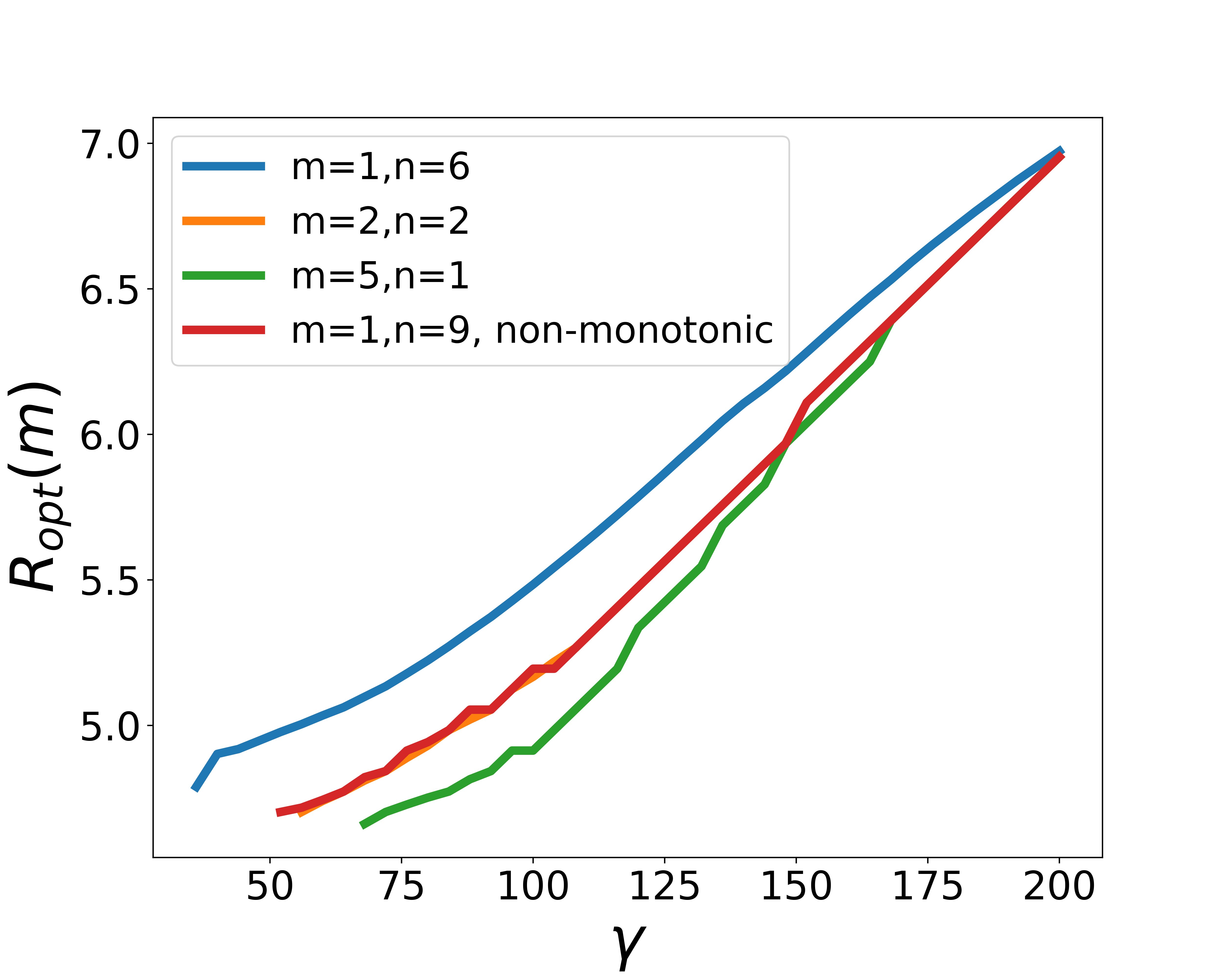}
}
\etbl
\caption{The current centre displacement as a function of $p_\|$ for $B_{z0}=-0.045T$ for different current profile shapes and the same total current $I_{RE}=200kA$.}
\label{fig:R(4)}
\end{figure*}

A more straightforward comparison of the current center displacement as a function of $p_\|$ and $B_{z0}$ for the four current profiles shown in Fig.\,\ref{fig:nshape} is plotted in Fig.\,\ref{fig:allPBO}. All cases are with the same constant total current $I_{RE}=200kA$. There the colour of the pseudo-colour map represent the major radius of the current center, with deep blue representing the cases where the equilibrium can not be found due to too significant current center displacement. All figures in Fig.\,\ref{fig:allPBO} shows the same trend that increasing the value of the vertical field (which points downward along $Z$ direction) pushes the current center inward, which is also expected for ordinary plasma equilibrium. Contrary to the ordinary equilibrium, changing the momentum of the runaway electrons without changing the total current or the current profile shape pushes the current center outward. Comparing Fig.\,\ref{fig:allPBO}(a), (b) and (c), it is found that the current center displacement occurs more easily for the more flattened current profile with the same amount of $p_\|$ change and $B_{z0}$ change, consistent with our observation above, and the ``boundary'' within which the equilibrium could be found becomes narrower for increasingly flattened current profile. Such trend can be more directly seen in Fig.\,\ref{fig:R(4)}, where the different colours represent different current profiles and $B_{z0}=-0.045T$. It is apparent that the peaked profile (blue) case takes more $p_\|$ change to achieve the same current center displacement compared with the flattened profiles (orange and green). Again, the non-monotonic current profile does not seem to have significant impact of the displacement susceptibility. 

Furthermore, from Eq.\,(\rfq{eq:FinalREGSEq}), it can be deduced that the deviation between the drift surfaces and the flux surfaces is determined by the relative strength of the second term of the RHS compared with that of the first term. One would thus naturally expect that with stronger runaway electron current, thus stronger poloidal magnetic field, the aforementioned deviation would decrease; while for smaller and more compact device, such deviation will be more pronounced. 

\begin{figure*}
\centering
\noindent
\btbl{cc}
\parbox{2.5in}{
    \includegraphics[scale=0.165]{renewpbo2e5_2_2.png}
}
&
\parbox{2.5in}{
	\includegraphics[scale=0.165]{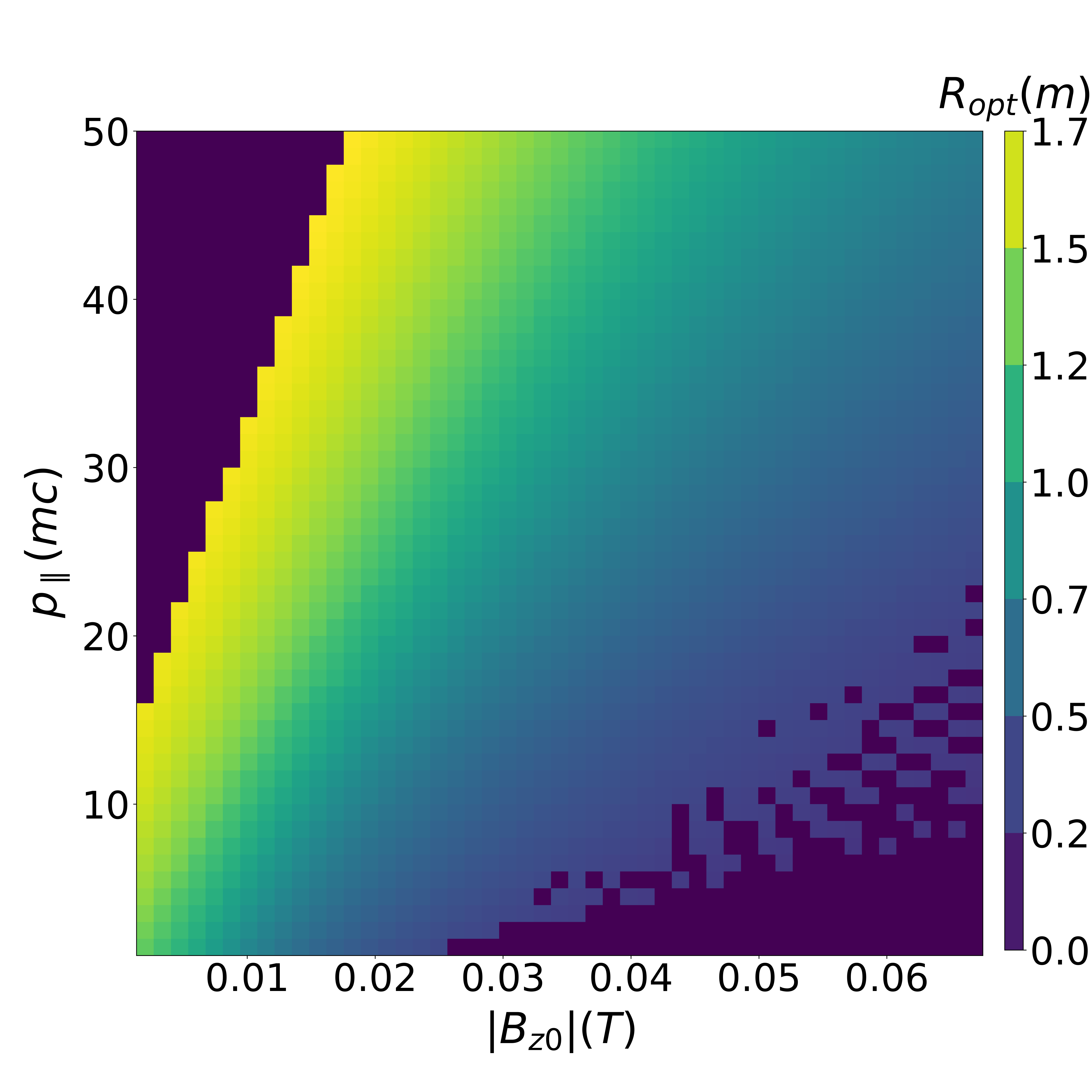}
}
\\
(a)&(b)
\etbl
\caption{The current center major radius position as a function of $p_\|$ and $B_{z0}$ for (a) the monotonic $m=2$, $n=2$ case, (b) the monotonic $m=2$, $n=2$ case. The two figures (a) and (b) depict the range of o point values under different devices, the (a) figure is taken from Fig.\,\ref{fig:allPBO} above when the o point value range is $m=2$, $n=2$  under the ITER device, and the (b) figure depicts the MAST, that is, the o point value range under the small device. All cases are with the same constant total current $I_{RE}=200kA$. The dark blue colour represent where the equilibrium cannot be found due to too much current displacement. }
\label{fig:MAST2_2}
\end{figure*}

To investigate this, we first compare two cases with the same $I_{RE}=200kA$ and monotonic current profile with $m=2$ and $n=2$, one with the same ITER-like geometry we have been using, the other with the MAST-like geomery as is described in Section \ref{ss:ANVFS}. The current center displacement as a function of their respective $p_\|$ and $B_{z0}$ are shown in Fig.\,\ref{fig:MAST2_2}, note the different range of the axes and the color bar. For the same amount of $p_\|$, the MAST-like case, with its smaller major radius, experience more significant horizontal drift compared with the ITER-like case. Mathematically, this is due to the $1/R$ contribution in the second term on the RHS of Eq.\,(\rfq{eq:FinalREGSEq}). Physically, this is due to the stronger centrifugal effect in the more compact device. So that we would expect the runaway electron drift behavior we show here to be more important for future small, compact machines.

\begin{figure*}
\centering
\noindent
\btbl{cc}
\parbox{2.5in}{
    \includegraphics[scale=0.225]{2e5newITER2-2_200-100.png}
}
&
\parbox{2.5in}{
	\includegraphics[scale=0.225]{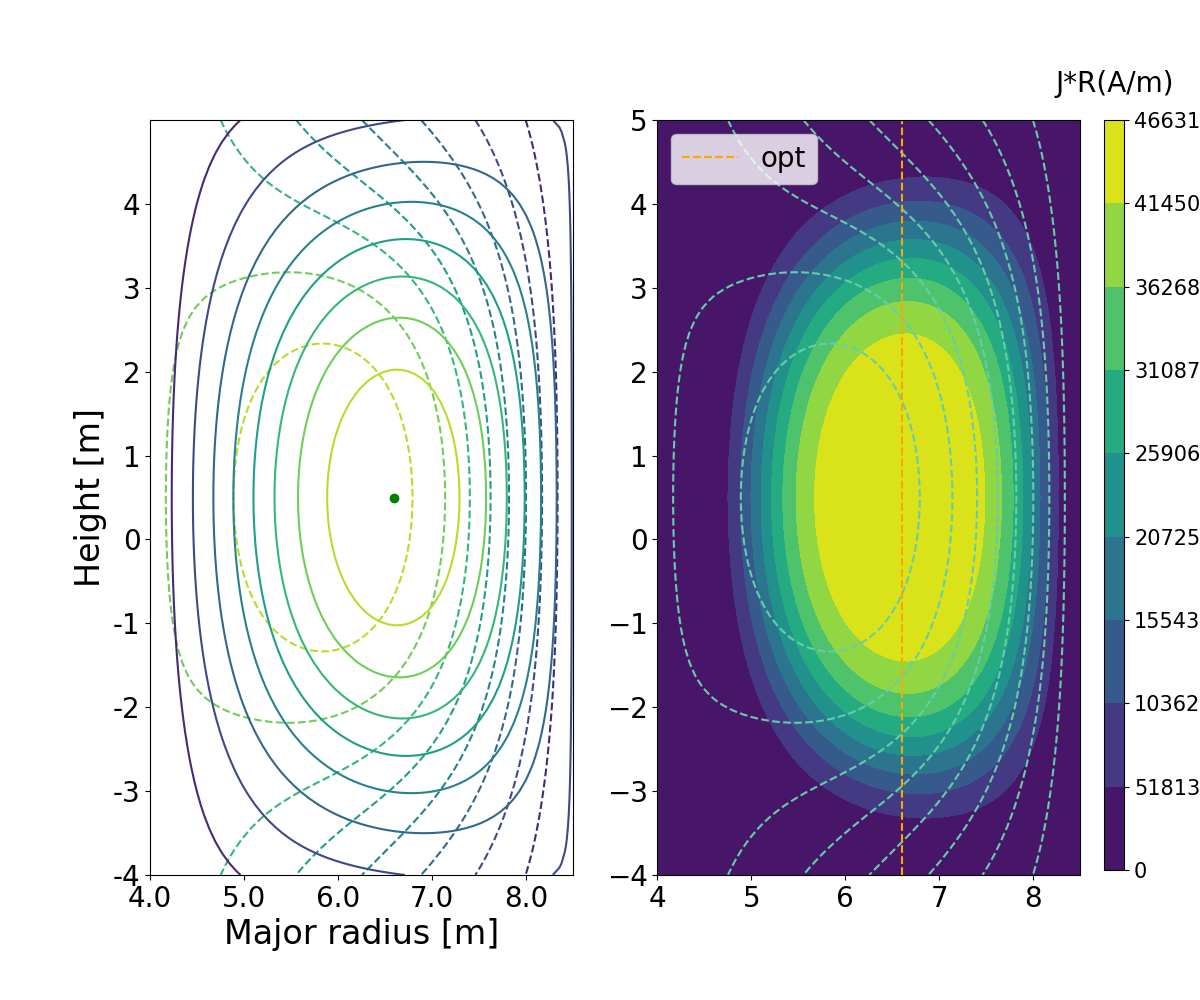}
}
\\
(a)&(b)
\\
\parbox{2.5in}{
    \includegraphics[scale=0.165]{renewpbo2e5_2_2.png}
}
&
\parbox{2.5in}{
	\includegraphics[scale=0.165]{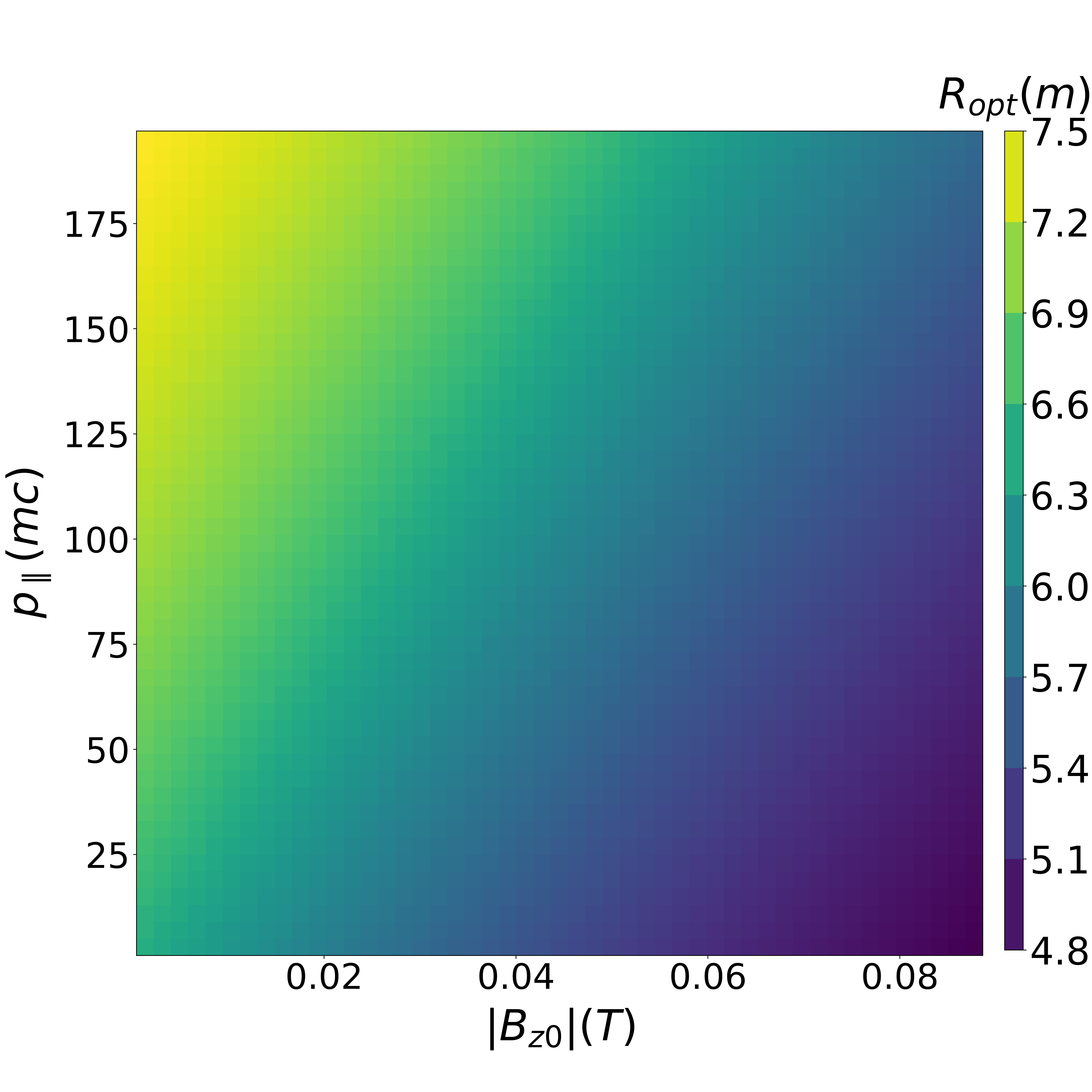}
}
\\
(c)&(d)
\etbl
\caption{(a) The equilibrium solution for $I_{RE}=200kA$ and (b) the equilibrium solution for $I_{RE}=1MA$. Both with $m=2$, $n=2$ monotonic current profile, $B_{z0}=-0.045T$, and  $\gg=100$. The current center major radius position as a function of $p_\|$ and $B_{z0}$ for both cases are shown in (c) and (d) respectively. Here (c) figure is taken from Fig.\,\ref{fig:allPBO} and corresponds to $I_{RE}=200kA$ and (d) corresponds to $I_{RE}=1MA$. Note the difference in the color bar range between (c) and (d).}
\label{fig:different1e6-2_2}
\end{figure*}

We then move on to investigate the effect of the runaway electron current with two ITER-like geometry cases, one with $I_{RE}=200kA$, the other with $I_{RE}=1MA$. Both cases share the same monotonic current profile shape with $m=2$ and $n=2$. Examples of their respective equilibrium, as well as the pseudo-colour plot of the current center displacement are shown in Fig.\,\ref{fig:different1e6-2_2}. Comparing Fig.\,\ref{fig:different1e6-2_2}(a) and (b), both cases using $p_\|=200mc$ and $B_{z0}=-0.045T$, it can be seen that the deviation between the drift surfaces and the flux surfaces is weaker for the $I_{RE}=1MA$ case and with less outward current center displacement, confirming our previous argument regarding such deviation depending on the relative strength of the $p_\|/qR$ term. Nevertheless, the runaway electron drift is still on the order of tens of centimeters for this high parallel momentum case, causing the runaway current equilibrium to deviate significantly from that of an ordinary equilibrium. Comparing Fig.\,\ref{fig:different1e6-2_2}(c) and (d), note the different color bar range, we could see that the current center displacement susceptibility is less for the $I_{RE}=1MA$ case for given amount of $p_\|$, consistent with our previous argument. Hence we would expect the above discussed runaway electron drift and the associated current displacement to be a significant effect mostly near the end of the runaway current depletion process when the total runaway current is small.

\subsection{Displacement of the guide center under asymmetric boundary conditions}
\label{ss:Asymmetric boundary conditions}

\begin{figure*}
\centering
\noindent
\btbl{cc}
\parbox{2.5in}{
    \includegraphics[scale=0.22]{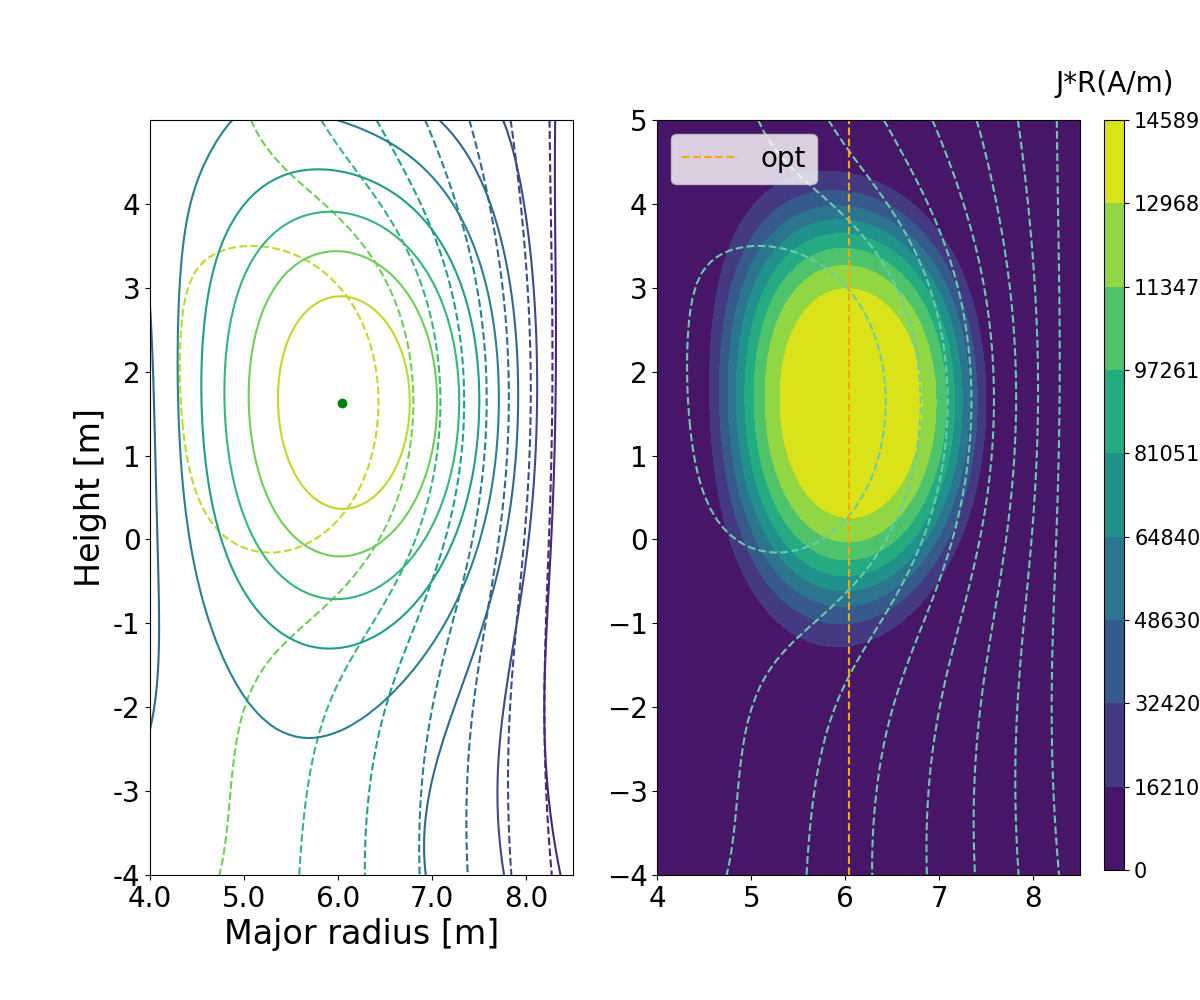}
}
&
\parbox{2.5in}{
	\includegraphics[scale=0.22]{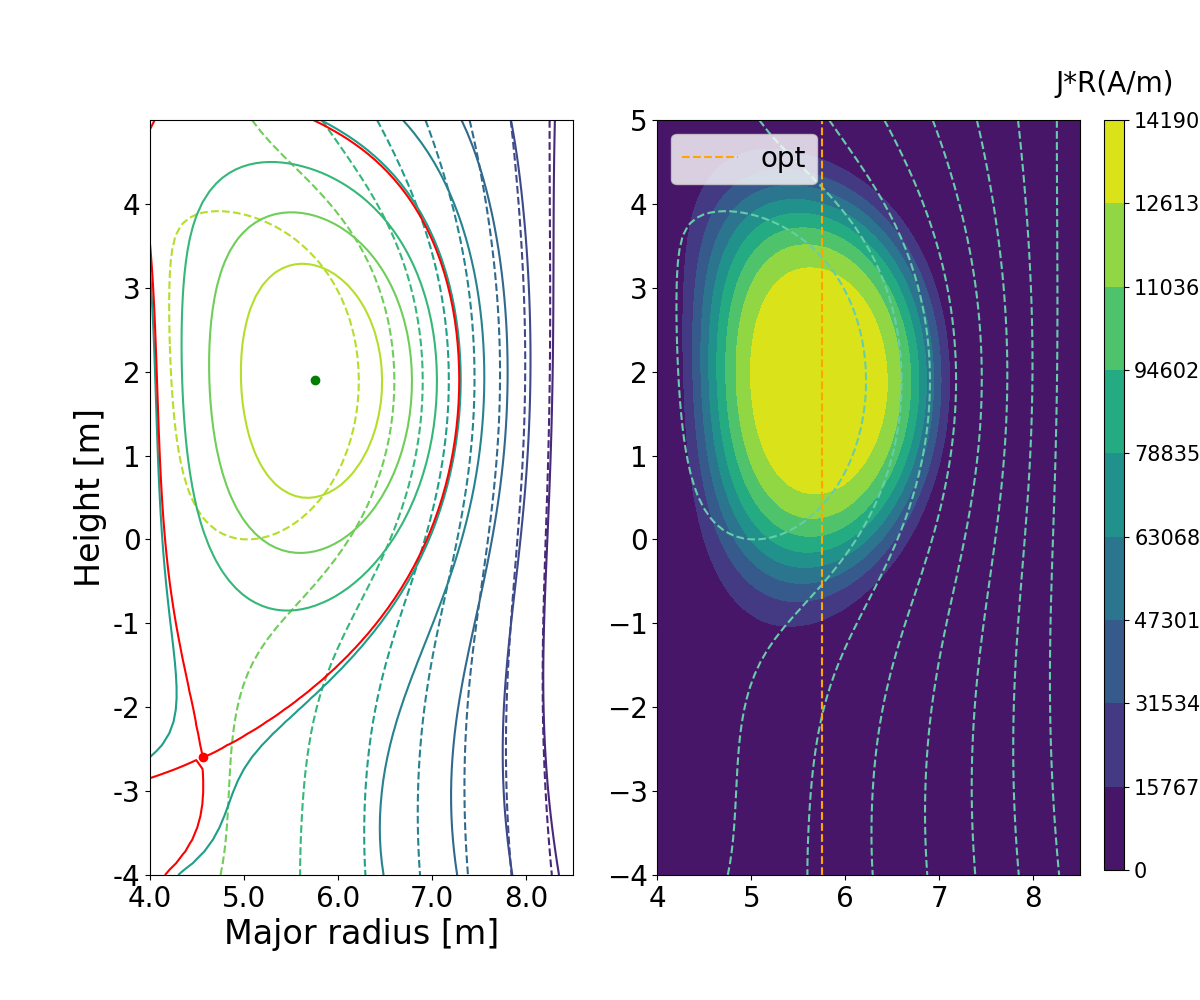}
}
\\
(a)&(b)
\\
\parbox{2.5in}{
    \includegraphics[scale=0.22]{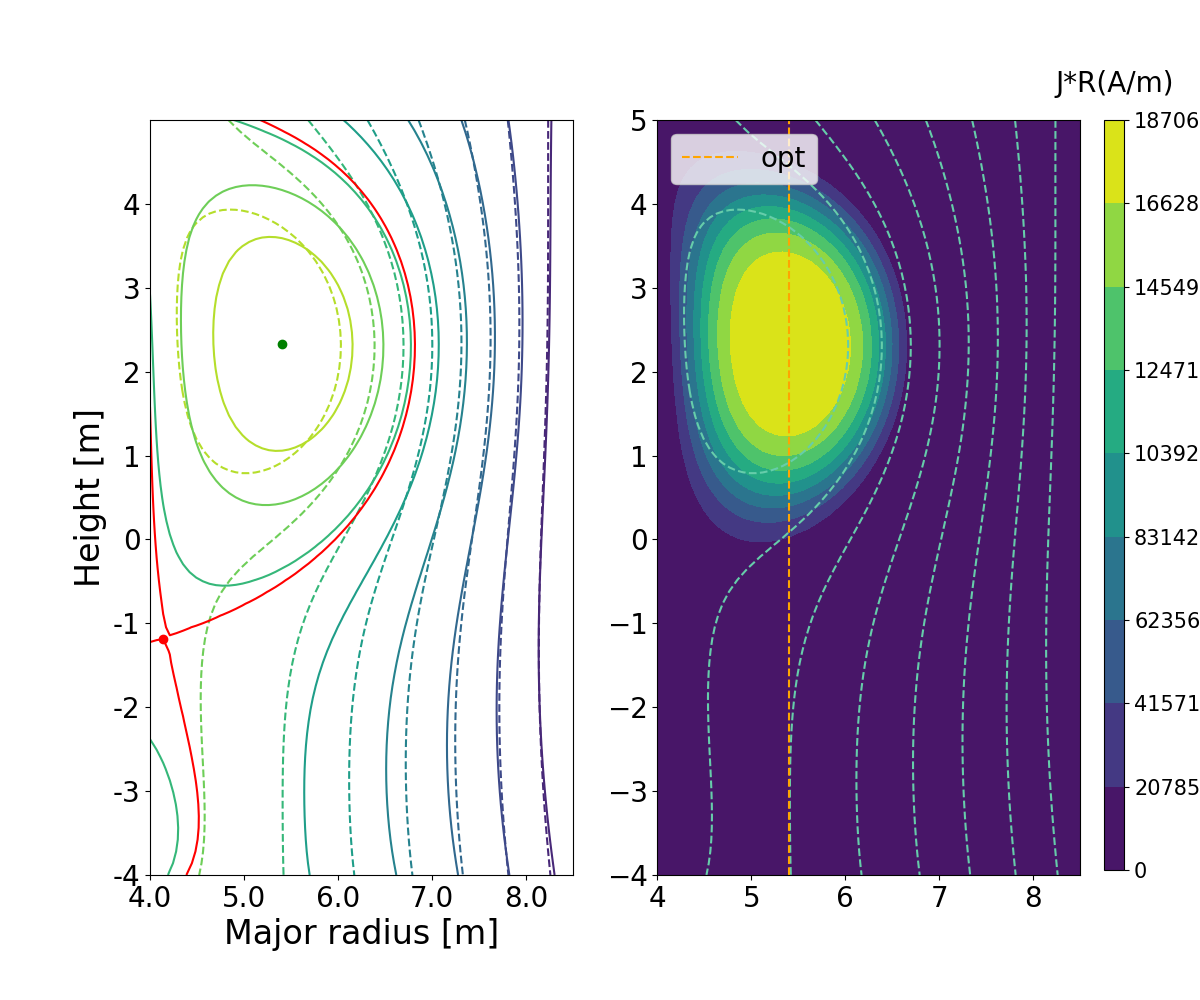}
}
&
\parbox{2.5in}{
    \includegraphics[scale=0.22]{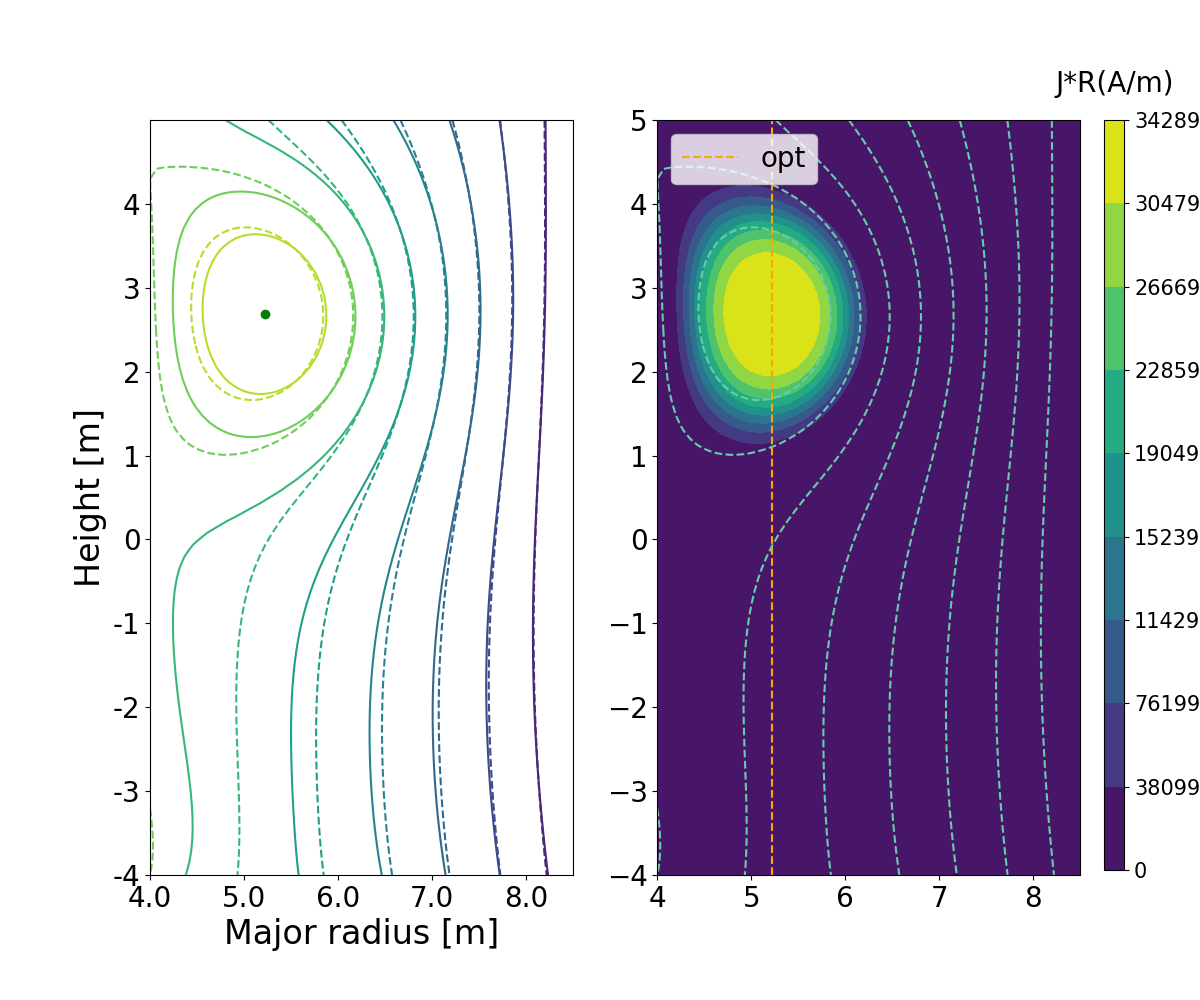}
}
\\
(c)&(d)
\etbl
\caption{The runaway current equilibrium with $B_{z0}=-0.018T$, and (a) $\gg=48$, (b) $\gg=36$, (c) $\gg=24$ and (d) $\gg=12$, respectively. The current profile is non-monotonic with $m=2$ and $n=2$ and total runaway electron current $I_{RE}=200kA$.  The position of the coil we add in this case is $x_u = 5m$, $x_d = 6m$, $y_u = 7.75m$, $y_d = -7.75m$. The magnitude of the current of the coil we add in this case is $I_u=85.94kA$, $I_d=77.35kA$. The solid contours represent the drift surfaces and the dashed contours represent the flux surfaces. The red dashed line represents the separatrix.}
\label{fig:dis2e5-2_2}
\end{figure*}

Last, let's take a look at the interesting case of an up-down asymmetric boundary condition, as is introduced in Section \ref{ss:ANVFS}. We set the upper coil current to be $I_u=85.94kA$ and the lower coil current to be $I_d=77.35kA$. The poloidal magnetic field generated by both coils would be reflected in the boundary condition. The equilibrium solutions under such boundary condition for several selected $p_\|$ and $B_{z0}=-0.018T$ are shown in Fig.\,\ref{fig:dis2e5-2_2}. Again, the total runaway electron current is $I_{RE}=200kA$ and the current profile is monotonic with $m=2$ and $n=2$.

From Fig.\,\ref{fig:dis2e5-2_2}(a) to (d), one could clearly see that as $p_\|$ decrease, there is an inward horizontal current displacement just as in the previous cases, but there is also an accompanied vertical displacement due to the asymmetry in the external poloidal field. Since it is rather common to have some up-down asymmetry in the poloidal field coil in real tokamaks, one would expect that the momentum change in real runaway current would also induce a corresponding vertical displacement during the current quench, which should be considered as an additional contribution to the vertical displacement in the ordinary equilibrium solution. Here we are satisfied with merely demonstrating the existence of such vertical displacement as a result of the runaway electron momentum change. The more detailed study with more realistic poloidal coil and 2D wall will be left for future works.

\section{Conclusion and discussion}
\label{s:Conclusion}

A Grad-Shafranov-like equation for the simple case that all current is carried by the runaway electrons and all runaway electron share the same parallel momentum is derived, solving directly for the runaway electron drift surfaces instead of the magnetic surfaces. A new force balance equation is obtained for the runaway electrons from their guide-center equation of motion, and combined with the toroidal symmetry to yield the aforementioned runaway current equilibrium equation, where an additional contribution from the runaway electron parallel momentum appears on the RHS. This additional contribution would result in the relative drift between the runaway electron drift surface and the flux surface. Since the runaway electron are the current carriers, such drift naturally result in the current channel displacement and thus the flux surfaces themselves are displaced, which in turn incurs further drift of the runaway electrons. With given runaway current profile shape and given boundary condition, we are able to numerically solve this new equilibrium equation. 

Numerical investigations are carried out using simple rectangular geometry with ITER-like and MAST-like geometric parameters. A number of different current profile shapes are tested. In general, it is found that with given external vertical field $B_{z0}$ and given total runaway electron current $I_{RE}$, the current center would exhibit a inward major radial displacement as the runaway electron momentum decreases, and like-wise an outward displacement as their momentum increases. The flattened current profile shows more susceptibility on this displacement compared with the peak profile, probably due to the more significant current scraping-off for the former. The hollowed current profile does not show significant impact on such current displacement susceptibility. As discussed above, this current displacement is caused by the relative drift between the runaway electrons and the flux surfaces, and its effect depends on both the relative strength of the parallel momentum to the poloidal field, and the major radius. Indeed, it is found that in smaller device the runaway electron drift is more apparent and the current center is more easily displaced by changes in the parallel momentum. Meanwhile, if the total current is large and the poloidal field is strong, such drift is milder, so is the current displacement. Thus we conclude that the above effect is more pronounced near the end of the runaway current depletion process when the total current is small, and so is the case in future compact devices. Furthermore, it is found that with up-down asymmetry in the boundary condition, such as those caused by asymmetric poloidal field coil in real tokamaks, the horizontal current displacement is accompanied by a vertical displacement. Hence the change in the runaway electron momentum could also additionally contribute to the vertical displacement event.

Despite the result obtained above, there is still a lot of room for future improvement of this new equilibrium theory. For instance, one could consider the situation where the parallel momentum is no longer a constant for all runaway electrons, but have a distribution in the configuration space instead. This would result in some additional term concerning $\na p_\|$, and is a natural future development of our theory. Moreover, one could also consider the scenario where the runaway electron distribution is not a $\gd$ function in the momentum space, but could have a distribution instead. To obtain the force balance equation in such a scenario, one have to construct the moment equations from the runaway electron kinetic equation to yield the equation of motion for the runaway electron fluid. Then one could combine such force balance equation and the symmetry of the system to obtain the Grad-Shafranov-like equation for the runway electron fluid just like we did in this paper. This is also one of the future works we will pursue. Another possible expansion of current work is to consider a more realistic 2D wall and the associated poloidal field reflecting that of a real tokamak instead of the uniform vertical field we used here. Yet another natural future expansion of the current simplified model is to extend to the higher parallel momentum regime of $p_\|/qBR\propto \ge$ instead of $p_\|/qBR\propto \ge^2$ in this paper. The curvature terms in Ref.\,\cite{Chang Liu2018NF} need to be included for this.

Nevertheless, the runaway current equilibrium theory presented in this study already provides a qualitative estimation on some of the important runaway current dynamic such as their relative drift to the flux surface and the current displacement as a result of parallel momentum change. Based on the equilibrium theory presented here and future works along similar lines, further stability analysis could be carried out for runaway electron dominated current and their consequential runaway electron transport or termination scenarios could be studied. These would help to achieve better runaway current control and depletion in the Current Quench phase of disruption.


\vskip1em
\centerline{\bf Acknowledgments}
\vskip1em

  The authors thank L.E. Zakharov for fruitful discussion. The authors thank Ben Dudson for the original open source equilibrium solver, upon which our numerical solver for the runaway current equilibrium is developed. This work is supported by the National MCF Energy R\&D Program of China under Grant No. 2019YFE03010001 and the National Natural Science Foundation of China under Grant No. 11905004.

\end{document}